\documentclass[12pt,a4paper,roman]{aa}  
\usepackage{rotating} 
\usepackage[english]{babel}
\usepackage{setspace}
\usepackage{siunitx}
\usepackage[utf8]{inputenc}
\usepackage{polski}
\usepackage[T1]{fontenc}
\usepackage{graphicx}
\usepackage{wrapfig}
\usepackage{makecell}
\usepackage{booktabs}
\usepackage{placeins}     % <-- add this in the preamble
\usepackage{txfonts}
\usepackage{amsmath}
\usepackage{float}
\usepackage{natbib}
\usepackage{graphicx}
\usepackage{enumitem}
\usepackage{dblfloatfix}
\usepackage{txfonts}
\usepackage{subcaption}
\usepackage{textcomp}
\usepackage{multirow}
\usepackage{caption}
\usepackage{tabularx}

\usepackage{amsmath}
\usepackage{float}
 \usepackage[table,xcdraw]{xcolor}
\usepackage[colorlinks=true, linkcolor=blue, citecolor=blue, urlcolor=blue]{hyperref}
\DeclareUnicodeCharacter{2212}{-}
\def \magperarcsec{mag arcsec$^{-2}$}
\def \mue{$\overline{\mu}_{\mathrm{eff},r}$}
\def \re{$r_{\mathrm{eff},r}$}
\def \msun{$\,M_{\odot}\,$}
\def \muo{$\mu_{0}$}
\begin{document}

   \title{From DES to KiDS: Domain adaptation for cross-survey detection of low-surface-brightness galaxies}

   \author{
    Hareesh Thuruthipilly \inst{1, 2, 3}
    \and Krzysztof Lisiecki \inst{1}
    \and Junais \inst{4, 5}
    \and Katarzyna ~Ma\l{}ek \inst{1}
    \and Agnieszka Pollo \inst{1}
    \and William J.~Pearson \inst{1}
    \and Antonio Vanzanella \inst{1}
    \and Saptarshi Pal \inst{1}
    \and Miguel Figueira \inst{1}
    \and Pratik Dabhade \inst{1}
    \and Anna Durkalec \inst{1}
    \and Aidan P. Cotter \inst{1}
    \and Unnikrishnan Sureshkumar \inst{1, 6}
    \and Nandini Hazra \inst{1}
    \and Patryk Matera \inst{1}
    \and Subhrata Dey \inst{1}
    \and Michal Vr\'abel \inst{1}
    \and Anirban Dutta \inst{1}
    \and Henry Willems \inst{1}
    \and Nicola Principi Cavaterra \inst{1} 
    \and Natalia Dobrowolska \inst{1}
    \and Wojciech Knop \inst{1}
    }

   \institute{National Centre for Nuclear Research, Pasteura 7, PL-02-093 Warsaw, Poland.
   \and 
   Departamento de Física Teórica, Atómica y Óptica, Universidad de Valladolid, 47011 Valladolid, Spain
   \and
Laboratory for Disruptive Interdisciplinary Science (LaDIS), Universidad de Valladolid, 47011 Valladolid, Spain \\
             \email{hareesh.thuruthipilly@uva.es}
              \and Instituto de Astrof\'{i}sica de Canarias, V\'{i}a L\'{a}ctea S/N, E-38205 La Laguna, Spain 
            \and Departamento de Astrof\'{i}sica, Universidad de La Laguna, E-38206 La Laguna, Spain
            \and  Wits Centre for Astrophysics, School of Physics, University of the Witwatersrand, Johannesburg, South Africa.}

   \date{Received XXX / Accepted YYY}
% \abstract{}{}{}{}{} 
% 5 {} token are mandatory
  \abstract
% context heading (optional)
{Low-surface-brightness galaxies (LSBGs) are vital for understanding galaxy formation, but their diffuse nature makes them challenging to detect. Upcoming large-scale surveys are expected to uncover a large number of LSBGs, which require robust automated methods to identify them across heterogeneous datasets. As a precursor to the Legacy Survey of Space and Time (LSST) and Euclid, we explore domain adaptation techniques to build homogeneous LSBG catalogues across current surveys.}
% aims heading (mandatory)
{We investigate the use of computer vision models and domain adaptation for cross-survey LSBG identification. Using models trained on the Dark Energy Survey (DES), we search for LSBGs in the Kilo-Degree Survey Data Release 5 (KiDS DR5). We then examine their structural and stellar population properties to pave the way for large-scale LSBG studies with LSST and Euclid.}
% methods heading (mandatory)
{We used an ensemble consisting of one convolutional neural network (CNN) and two transformer models trained on DES cutouts and applied to KiDS DR5 imaging with surface-brightness normalisation. Structural parameters were estimated with {\tt galfitm}, and the sample was further refined through visual inspection to produce the final candidate sample. Photometric redshift and stellar population properties were estimated through spectral energy distribution (SED) fitting with CIGALE.}
% results heading (mandatory)
{We identify 20\,180 LSBGs and 434 Ultra-diffuse galaxies (UDGs) in KiDS DR5. Their structural parameters are similar to the known LSBGs from DES and Hyper Suprime-Cam SSP Survey (HSC-SSP). The KiDS-LSBGs follow a continuous size–luminosity relation connecting classical dwarf galaxies and UDGs, and their colours are bimodal ($\sim$73\% blue, $\sim$27\% red).
Cross-matching with spectroscopic and cluster catalogues provides redshifts for 4\,913 systems, enabling a systematic characterisation of the star-forming main sequence of LSBGs. Photometric redshifts derived via SED fitting are mildly overestimated ($\simeq 0.024$), leading to systematic offsets in stellar mass and star formation rate estimates. However, these biases induce only small shifts ($\sim$0.13-0.22 dex) in specific star formation rate, thereby preserving the structure of the star-forming main sequence. Strong environmental trends are also evident, with cluster LSBGs and UDGs exhibiting redder colours and reduced star formation compared to non-cluster systems. This indicates efficient quenching driven by environmental processes.}
% conclusions heading (optional)
{We demonstrate that, with domain adaptation, cross-survey LSBG identification can be achieved with deep learning models. Thus, the methodology presented here provides a powerful and scalable pathway for constructing homogeneous LSBG catalogues across surveys. This framework is well-suited for the era of LSST and Euclid, where millions of diffuse galaxies will be discovered, and consistent cross-survey classification will become essential.}


\titlerunning{KiDS-LSBGs.} 
   \maketitle

\section{Introduction}\label{introduction}
Low-surface-brightness galaxies (LSBGs) are typically defined as galaxies with central surface brightness fainter than the night sky \citep{Bothun_1997}. Despite their diffuse nature, which makes them challenging to detect, both simulations \citep[e.g.][]{Martin_2019, Wright_2025_romulus} and observations \citep[e.g.][]{Dalcanton_1997, Neil_2000} predict that the bulk of the galaxy population resides in the low-surface-brightness regime. For instance, LSBGs are expected to account for between 30\% and 60\% of the total galaxy number density \citep{McGaugh_1996, Haberzettl_2007, Martin_2019, Wright_2025_romulus} and up to $\sim$15\% of the dynamical mass content of the Universe \citep{Driver, Minchin}. Nevertheless, LSBGs remain severely underrepresented in current galaxy catalogues, making them an important yet poorly understood component of the galaxy population \citep{Kaviraj_2020}.

However, despite their significance, there is no single, universally adopted observational definition of LSBGs in the literature. For example, several studies identify LSBGs using mean surface brightness thresholds in the optical bands, such as $\overline{\mu}_{\mathrm{eff},g} > 24.2$–$24.3 \ \mathrm{mag \ arcsec^{-2}}$ (e.g., \citealt{Greco, Tanoglidis1}) or \mue{} $> 23 \ \mathrm{mag \ arcsec^{-2}}$. Other works instead prefer definitions based on the central surface brightness, typically using criteria like \muo{}$_{,B} > 22.0-22.5 \ \mathrm{mag \ arcsec^{-2}}$ (e.g., \citealt{Christie25} and \citealt{Du_2015}, respectively). In addition, for imaging-only data, angular size cuts (e.g., $r_{\mathrm{eff}} > 2.5\arcsec$) are frequently applied to reduce contamination from distant compact sources \citep{Greco, Tanoglidis1}. This observational ambiguity likely reflects a physically diverse population, suggesting that multiple formation pathways contribute to the observed properties of LSBGs.

From a theoretical standpoint, simulations suggest that \mbox{LSBGs}
 can form through both intrinsic and extrinsic mechanisms. Intrinsically, LSBGs may share progenitors with high-surface-brightness galaxies (HSBGs). In this scenario, strong early supernova feedback can produce diffuse stellar distributions \citep{Martin_2019}, whereas high halo and stellar spin can prevent central collapse \citep{Perez_2022}. Extrinsically, LSBGs can arise from co-planar, co-rotating mergers or aligned gas accretion, which increase angular momentum and suppress efficient star formation \citep{Cintio_2019, Wright_2025_romulus}.

These formation channels suggest that LSBGs constitute a highly diverse population with several subclasses, including the even more diffuse and observationally challenging ultra-diffuse galaxies (UDGs). UDGs have effective radii comparable to Milky Way–like galaxies ($r_{\mathrm{eff},r} \gtrsim 1.5~\mathrm{kpc}$) but much lower stellar masses ($10^{6}$–$10^{9}~M_{\odot}$).
Similarly to LSBGs, their observational definitions vary in the literature, typically based on either central $g$-band \citep{vanDokkum} or mean $r$-band \citep{Yagi, Burg} surface brightness thresholds. Moreover, the existing criteria in the literature do not take into account dust correction, which can significantly move UDG candidates outside standard surface brightness-size criteria (see  Vanzanella et al. in prep.). Another extreme subclass is giant LSBGs, which are faint, highly extended, and extremely gas-rich ($M_{\rm HI} > 10^{10}$ \msun{}; \citealt{Sprayberry1995, Junais_malin2}). The origins of these extreme systems remain debated, providing testing platforms for galaxy formation and cosmological models \citep{Cintio_2017, Laudato, Montes}.

With the advancements in deep imaging from large-scale surveys such as the Dark Energy Survey (DES; \citealt{DESDR2}), the Kilo-Degree Survey (KiDS; \citealt{Wright_2025_kids_dr5}), the Hyper Suprime-Cam Subaru Strategic Program (HSC-SSP; \citealt{Aihara}), Euclid \citep{laureijs2011euclid}, and the Vera C. Rubin Observatory’s Legacy Survey of Space and Time (LSST; \citealt{Ivezi__2019}), more and more LSBGs are being discovered \citep{Greco, Tanoglidis1, haressh_lsbs, Marleau_2025_EDR1, Marleau_2025_Q1, Su_2025_kiDS_udgs, Venhola_2026}. One of the common challenges in the search for LSBGs is the presence of contaminants such as diffuse light from bright galaxies, galactic cirrus, spiral-arm tails, and tidal streams \citep{Tanoglidis1, Du_2024}. Semi-automated approaches to remove these contaminants often yield low success rates and require labour-intensive visual inspections.

Recent advances in deep learning (DL) offer promising solutions to the above-mentioned problem, with convolutional neural networks (CNNs) and transformers now widely applied to both observational and simulated astronomical data (e.g. \citealt{Hareesh, Hareesh2, Jia, Grespan, Margalef_2024}). LSBG searches in large surveys such as HSC-SSP and DES have enabled the construction of training sets \citep{Greco, Tanoglidis1} for DL models, leading to successful applications of DL techniques for LSBG identification \citep{Tanoglidis2, Yi_2022, Xing2023, Su_LSBGNET, haressh_lsbs}.
However, most of these studies test their models and analyse data from a single survey. They do not evaluate how well models trained on one survey perform when applied to another.

Recently, this problem has been explored by \citet{Su_LSBGNET} using transfer learning. They trained a DL model on Sloan Digital Sky Survey (SDSS) images, and tested it on DES data, achieving a recall\footnote{Recall is the fraction of true positive cases that are correctly identified by the model.} of $\sim78\%$, which increased to 97\% after retraining with DES images. However, retraining with manually labelled data is not always feasible. A cross-survey methodology that can find LSBGs across surveys without the need for retraining is necessary in the era of LSST and Euclid. 

In this context, in \citet{hareesh_2025_hsc}, we demonstrated that, by using a surface-brightness-based normalisation, a model trained on one survey (DES) can be directly applied to a different, even deeper survey (HSC) without retraining and achieve a recall as high as 93\%. This approach can be understood as a form of domain adaptation, where the goal is to bridge the distributional differences between the source domain (on which the model is trained) and the target domain (to which the model is applied). Domain adaptation also comes under the broader umbrella of transfer learning (see \citealt{Kouw2019ARO} and \citealt{Farahani_2021_domain_adaptation} for a detailed discussion on domain adaptation) and is being increasingly used for cross-survey applications \citep{Ciprijanovic_2023, Ye_2025, Treyer_2026} and simulation-to-observation transfer \citep{Belfiore_2025, Parul_2025, Lopez_Cano_2026}.

In this work, we investigate whether the methodology presented in \citet{hareesh_2025_hsc} can be further expanded to large-scale sky surveys. We use the labelled dataset of LSBGs and contaminants identified in DES by \citet{Tanoglidis1} and updated by \citet{haressh_lsbs} to train DL models. These models are then applied to the KiDS Data Release 5 (KiDS DR5) to identify LSBGs. 
Here, we classify LSBGs as sources that satisfy the surface brightness criterion $\overline{\mu}_{\mathrm{eff},r} > 23.0~\mathrm{mag~arcsec^{-2}}$ and have a minimum non-circularised $r$-band half-light radius of $r_{\mathrm{eff},r} > 2.5\arcsec$. For UDGs, we follow \citet{Burg,vanderBurg2017} and define UDGs as galaxies with $\overline{\mu}_{\mathrm{eff},r} > 24.0~\mathrm{mag~arcsec^{-2}}$ and $r_{\mathrm{eff},r} > 1.5~\mathrm{kpc}$. Here, the $r$ band is adopted instead of the $g$ or $B$ bands due to its greater sensitivity to older stellar populations.

 The paper is organised as follows: Sect. \ref{data} discusses the data, Sect. \ref{method} provides a brief overview of the methodology, including the model architecture, training, and visual inspection. The results are presented in Sect. \ref{results}, followed by a discussion of the results and the properties of the newly identified LSBGs in Sect. \ref{discs}. Finally, Sect. \ref{conclusion} summarises the conclusions of our analysis. Throughout this paper, we assume a flat $\Lambda$CDM cosmology with parameters $(H_0, \Omega_{\rm M}, \Omega_\Lambda) = (70~\mathrm{km~s^{-1}~Mpc^{-1}}, 0.3, 0.7)$.
\section{Data}\label{data}
 
\subsection{Labelled LSBGs and contaminants from DES}\label{traing_des_data}
The DES observed $\sim5000 \text{ deg}^2$ of the southern Galactic cap using the Dark Energy Camera (DECam). It used \textit{g,r, i,z, Y} photometric bands with approximately 10 overlapping dithered exposures in each filter (90 sec in \textit{griz}-bands and 45 sec in the \textit{Y}-band).  The DES Data Release 1 (DES~DR1) delivers a median point-spread function (PSF) full width at half maximum (FWHM) of $1.12\arcsec $ in the $g$ band, $0.96\arcsec $ in the $r$ band, and $0.88\arcsec $ in the $i$ band.
It also achieves a photometric precision better than 1\% in all bands and an astrometric precision of $\sim151$~mas \citep{DR1}.
For DES DR1, \citet{Tanoglidis1} estimated the median $3\sigma$ surface-brightness depth within a $10\arcsec \times 10\arcsec$ region to be $28.26^{+0.09}_{-0.13}$, $27.86^{+0.10}_{-0.15}$, and $27.37^{+0.10}_{-0.13}~\mathrm{mag\,arcsec^{-2}}$ in the $g$, $r$, and $i$ bands, respectively. The quoted uncertainties correspond to the 84th and 16th percentiles of the depth distribution across DES tiles.

For training, validation, and testing of the models, we used the labelled dataset of LSBGs and contaminants identified from DES DR1 by \citet{Tanoglidis1} and extended by \citet{haressh_lsbs}. In \citet{haressh_lsbs}, it was shown that some of the 20\,000 contaminants from DES listed in the publicly available catalogue\footnote{\url{https://github.com/dtanoglidis/DeepShadows/blob/main/Datasets}}
 were not contaminants but genuine LSBGs. 
After removing these sources, 18\,468 contaminants remained in the list, all of which were labelled as Class 0 (contaminants). To maintain an approximately balanced dataset, we randomly selected 18\,532 LSBGs from the extended DES sample and assigned them a label of Class 1, resulting in a total of 37\,000 objects. Although the class distribution is not perfectly equal, small deviations from class balance have been shown not to significantly affect model performance \citep{Buda_2018}.

We generated cutouts for each object using DES DR1 images in the \textit{g}-band and \textit{r}-band. Each cut-out covers a $40\arcsec \times 40\arcsec$ region of the sky (equivalent to $152 \times 152$ pixels) and is centred on the coordinates of the object (either an LSBG or a contaminant). Here, the cutout size was chosen based on the size distribution of known DES LSBGs from \citet{Tanoglidis1} and \citet{haressh_lsbs}, which are, by selection, constrained to have half-light radius $< 20\arcsec$. Following \citet{hareesh_2025_hsc}, the pixel values of each cut-out were converted to surface-brightness units ($\mu\mathrm{Jy\,arcsec^{-2}}$). The cutouts were then resized to $64 \times 64$ pixels using the {\tt skimage.transform}\footnote{\url{https://scikit-image.org/docs/stable/api/skimage.transform.html}} Python package to reduce the computational cost. 
The resized cutouts cover the same sky area as the originals, but with the DES pixel scale changed from $0.263$ to $0.625~\mathrm{arcsec\,pix^{-1}}$. The resized $g$- and $r$-band cutouts are stacked, in that order, to form two-channel inputs for training. Although KiDS also provides $i$-band data, we use only the $g$ and $r$ bands to reduce computational costs. In addition, previous studies have shown that models trained using two bands \citep{hareesh_2025_hsc} achieve performance comparable to three-band models \citep{haressh_lsbs}, with similar accuracies ($\simeq 95\%$).

Our training catalogue contains 37\,000 objects, comprising 18\,532 LSBGs and 18\,468 contaminants. Before training, we randomly split the full training sample into a training set (32\,000 objects; 87\%), a validation set (2\,000 objects; 5\%), and a test set (3\,000 objects; 8\%). Each of these subsets maintained similar proportions of LSBGs and contaminants to ensure balanced representation. 
Examples of LSBGs and contaminants in the training set are shown in Fig. \ref{fig:training}.

\subsection{Kilo-Degree Survey}
KiDS is an optical wide-field imaging survey conducted with the VLT Survey Telescope (VST) using the OmegaCAM camera at the European Southern Observatory’s Cerro Paranal Observatory in Chile. The KiDS footprint is also covered in the near-infrared by the VISTA Kilo-Degree Infrared Galaxy Survey (VIKING; \citealt{Edge_2013}) in the \textit{ZYJHK} bands. Together, the combined KiDS+VIKING dataset provides a nine-band optical–NIR coverage, offering the broadest wavelength range among the Stage III imaging surveys. This makes it particularly powerful for obtaining accurate photometric redshifts (photo-$z$) and for galaxy evolution analyses.

The final public data release (DR5) covers $\sim 1350\, \text{deg}^2$ in four optical filters (\textit{ugri})\footnote{In KiDS DR5, the $i$-band data were collected by the survey in two phases: $i_1$ (partial coverage) and $i_2$ (main homogeneous coverage). In this work, we use $i_2$ and refer to it as $i$.}, reaching a median $r$-band $5\sigma$ limiting magnitude of 24.8 mag \citep{Wright_2025_kids_dr5}. The DR5 includes reduced optical imaging, single-band detection catalogues, and multi-band catalogues that incorporate the matched VIKING photometry. All data products are available through the European Southern Observatory (ESO) archive. 

For this work, we retrieved the imaging data in the \textit{g}, \textit{r}, and \textit{i} bands, along with the corresponding multi-band catalogues. The dataset consists of 1347 tiles per band, each corresponding to an individual KiDS pointing.
We measured the median $3\sigma$ surface-brightness limiting depths to be $\mu_{3\sigma,g}=28.86\pm0.08$, $\mu_{3\sigma,r}=28.74\pm0.1$, and $\mu_{3\sigma,i}=27.44\pm0.19~\mathrm{mag\,arcsec^{-2}}$. The quoted uncertainties represent the median absolute deviations (MAD) across the full set of tiles. The distributions of the $3\sigma$ surface-brightness depths in the \textit{g}, \textit{r}, and \textit{i} bands for the KiDS tiles are shown in Figures~\ref{kids_g_depth_map}, \ref{kids_r_depth_map}, and \ref{kids_i_depth_map}, respectively. Full details of the depth estimation procedure are provided in Appendix~\ref{appendix:surface_brightness_depth}.

\subsection{Pre-selected LSBG candidate sources}\label{catalog_preselection}
The KiDS DR5 multi-band catalogue, adopted as the parent sample for identifying LSBGs, contains approximately $1.38 \times 10^{8}$ sources.
To select LSBGs and minimise contamination, we applied additional cuts on source size, shape, surface brightness, and colour following \citet{Greco} and \citet{Tanoglidis1}.
We restricted the effective radius to the range $2\arcsec$–$20\arcsec$, based on the \texttt{FLUX\_RADIUS}. Although the LSBG definition adopts \re{}$>2.5\arcsec$, we relaxed the lower limit to $2\arcsec$ to account for the systematic underestimation of effective radii by {\tt SExtractor}, which is typically $\sim20\%$ \citep{hareesh_2025_hsc, Su_2025_kiDS_udgs}.
In addition, we required an axis ratio of $q = {\tt B\_IMAGE}/{\tt A\_IMAGE} > 0.3$ to remove artefacts such as diffraction spikes.
The effective surface brightness in the $r$ band was computed as
\begin{equation}\label{msb_relation}
\overline{\mu}_{\mathrm{eff},r} = m_r + 2.5 \log_{10} (2\pi r_{\mathrm{eff},r}^2 q),
\end{equation}
where $m_r$ is the total $r$-band magnitude from {\tt MAG\_AUTO}, $r_{\mathrm{eff},r}$ is the effective radius, and $q$ is the axis ratio. The resulting values were constrained to 23 < \mue{} < 28.8~\magperarcsec{}, with the upper limit set by the $\mu_{3\sigma,r}$. To further reduce contamination from optical artefacts, blends, and
high-redshift galaxies, we applied the colour selections of
\citet{Greco, Tanoglidis1} using the KiDS Gaussian Aperture and PSF (GAaP) photometry\footnote{GAaP photometry uses PSF-matched Gaussian apertures to obtain consistent multi-band fluxes and colours.}, which is reported only as colours between adjacent bands. Since KiDS~DR5
does not provide a direct ${\tt COLOUR\_GAAP\_g\_i}$ colour, we
constructed it from the available colours as
${\tt COLOUR\_GAAP\_g\_i}
= {\tt COLOUR\_GAAP\_g\_r}
  + {\tt COLOUR\_GAAP\_r\_i}$. The selection criteria were then applied in terms of these GAaP colours:
\begin{align*}
\centering
-0.1 &< {\tt COLOUR\_GAAP\_g\_i} < 1.4, \\
{\tt COLOUR\_GAAP\_g\_r}
&> 0.7\times{\tt COLOUR\_GAAP\_g\_i} - 0.4, \\
{\tt COLOUR\_GAAP\_g\_r}
&< 0.7\times{\tt COLOUR\_GAAP\_g\_i} + 0.4.
\end{align*}
The colour–colour diagram of the full KiDS DR5 parent sample, along with the region (indicated by the black box) used to preselect LSBG candidates, is shown in Fig.~\ref{fig:color_preselctions}. The adopted colour preselection is consistent with the colours reported for LSBGs in the literature \citep{Greco, Tanoglidis1, Zaritsky_SMUDGE, Su_LSBGNET}.

We note that the DES training sample defines LSBGs using a $g$-band based threshold ($\overline{\mu}_{\mathrm{eff},g} > 24.2~\mathrm{mag~arcsec^{-2}}$), whereas our target selection in KiDS adopts an $r$-band criterion ($\overline{\mu}_{\mathrm{eff},r} > 23.0~\mathrm{mag~arcsec^{-2}}$). This adjusted threshold accounts for the typical $g-r$ colours of LSBGs ($\sim$0.0–1.0 mag; \citealt{Greco, Tanoglidis1}), ensuring that the two selections probe broadly similar populations. The $r$-band choice also enables us to quantify the impact of bandpass selection and any resulting biases on the recovered LSBG sample.
This will also allow us to assess how existing LSBG catalogues can be leveraged to streamline discovery in the significantly deeper and wider surveys such as LSST and Euclid.
After applying these criteria, the resulting catalogue contained 322\,278 candidate sources. Follow-up selection procedures (Sect. \ref{method}) are then applied to these sources.

\section{Methodology} \label{method}
% \subsection{Computer Vision Models}\label{CVM}
\subsection{Convolutional Neural Networks}\label{cnn}
A convolutional neural network (CNN) uses multiple stacked convolutional layers to progressively learn hierarchical features, enabling the recognition of patterns at increasing levels of abstraction  \citep{Mallat_2016}.
In this work, we construct a CNN based on the architectures of \citet{Schaefer_2018} and \citet{Hareesh}. The model consists of four convolutional blocks with 16, 32, 64, and 128 filters (all with $3\times3$ kernels). Each block contains two convolutional layers, with batch normalisation (BN) layers applied after each layer. In addition, all the convolution layers in blocks 3 and 4 have dropout (rate = 0.6) layers after the BN layers. The first two blocks additionally include max-pooling layers at the end of the block. All convolutional layers employ L2 weight regularisation. The extracted feature maps are flattened and fed into three fully connected layers (1024, 512, and 128 units). All the layers, including the CNN layers, use the Exponential Linear Unit (ELU) activation. A single sigmoid unit is used in the final layer for binary classification, and the network is trained using the Adam optimiser \citep{kingma2017adam} with a binary cross-entropy loss. Training employs an early-stopping scheme \citep{Prechelt1996EarlySW}, in which training stops if the validation loss does not reduce for 20 (patience) consecutive epochs. The trained CNN model is called the LSBG CNN from hereafter, and the design of the model is shown in Fig. \ref{fig:LSBG_CNN}. During training, the input data were augmented using random horizontal and vertical flips. No additional normalisation was applied to the input data during either training or inference.

\subsection{Transformer Models}\label{transformer}
Transformers are DL architectures that employ self-attention mechanisms to capture correlations within input features, making them powerful for prediction and classification tasks \citep{vaswani2017attention}. In this work, we employ two transformer-based models: the LSBG Detection Transformer (LSBG-DETR) and the LSBG Vision Transformer (LSBG-ViT).

These models were originally developed and applied to identify LSBGs in DES~DR1 by \citet{haressh_lsbs}. Unlike the original study, which employed ensembles of each architecture with three-band input images, we adopt only a single model from each architecture (LSBG-DETR~1 and LSBG-ViT~2 from \citealt{haressh_lsbs}) and retrain them using two-band inputs for application to the KiDS dataset. This approach reduces the computational cost, while the combination of the LSBG-DETR, LSBG-ViT, and CNN models still provides an effective ensemble at the cross-model level.

The design of the LSBG-DETR and LSBG-VIT models is shown in Fig. \ref{fig:lsbg_detr} and Fig. \ref{fig:lsbg_vit}, respectively.
Further details of the model architectures and training procedures are provided in \citet{haressh_lsbs}. Both models were trained using the Adam optimiser with a binary cross-entropy loss function and a batch size of 128. An exponentially decaying learning rate was adopted, with an initial value of $10^{-4}$ and a decay rate of $0.9$ applied every $10^4$ training steps, where one step corresponds to a single batch update. Early stopping was employed with a patience of 20 epochs. During training, the input data were augmented using random horizontal and vertical flips. No additional normalisation was applied to the input data during either training or inference. 

\subsection{Domain adaptation for cross-survey LSBG search}\label{transfer_learning}
In this work, we use models trained for LSBG classification on DES to find LSBGs in KiDS. Although the task remains the same, differences in the datasets place this approach under the regime of domain adaptation, a subfield of transfer learning. In general, several challenges arise when applying a model trained on one survey to another: (i) differences in photometric zero-points, (ii) variations in pixel scale, (iii) differences in the PSF, (iv) variations in data depth and noise properties, and (v) survey-specific instrumental artefacts. Differences in filter transmission curves may also introduce minor systematic variations; however, these typically have a smaller impact on the apparent shape and size of galaxies, and therefore only weakly affect their identification as LSBGs.

Currently, survey-specific instrumental artefacts cannot be removed using a single generalised approach, as they are intrinsically dependent on the properties of individual surveys.  However, \citet{hareesh_2025_hsc} demonstrated that converting imaging data to physical surface-brightness units ($\mu\mathrm{Jy\,arcsec^{-2}}$) effectively accounts for zero-point differences between surveys. Similarly, resizing images (while conserving flux) to a common lower pixel scale allows models to generalise across datasets. Since the PSFs of the ground-based imaging surveys considered here are broadly comparable, the impact of PSF differences on LSBG classification is expected to be negligible. With these normalisations, \citet{hareesh_2025_hsc} demonstrated that models can successfully identify sources represented in the training dataset, even when applied to deeper data. 

Hence, the methodology described in \citet{hareesh_2025_hsc} is a physics-motivated form of unsupervised domain adaptation. By converting pixel values to surface-brightness units ($\mu\mathrm{Jy,arcsec^{-2}}$) and resizing the images to a common resolution, we directly align the physical input distributions of LSBGs in KiDS data with those in DES data. This enables robust cross-survey LSBG identification without any retraining. A key limitation, however, is that model recall drops in the very faint regime ($g \geq 22$), which is under-represented in the training data. Since KiDS and DES have comparable depths, this is unlikely to affect this work, but it should be addressed for surveys such as LSST and Euclid. 

Following this approach, we adopt the DES-trained models (described in Sect.~\ref{cnn} and Sect.~\ref{transformer}) to identify LSBGs in KiDS. For the selected sources from the KiDS DR5 parent catalogue (Sect.~\ref{catalog_preselection}), we produced \textit{g}- and \textit{r}-band cutouts. Each cutout covers a $40\arcsec \times 40\arcsec$ region centred on the source. Given the KiDS pixel scale of $0.2\arcsec$ per pixel \citep{Wright_2025_kids_dr5}, this corresponds to $200 \times 200$ pixels, consistent with the DES cutouts. 
The pixel values are converted to surface-brightness units ($\mu\mathrm{Jy\,arcsec^{-2}}$). The cutouts are then rescaled to $64 \times 64$ pixels using the \texttt{skimage.transform} package to match the input size required by the CNN and transformer models. The rescaled \textit{g} and \textit{r} images were then stacked and passed to the models described in Sect.~\ref{cnn} and Sect.~\ref{transformer}. To maximise completeness, we classified a source as an LSBG candidate if at least one model returned a probability greater than 0.5. While this threshold increases sensitivity to faint systems, it can also introduce contamination that requires subsequent cleaning.
Applying this criterion yielded 28\,971 candidate LSBGs.

\subsection{S\'ersic fitting}\label{sersic_fitting}
We used {\tt galfitm}, the multi-band extension of {\tt galfit} developed in the context of the MegaMorph project \citep{galfit, Barden_2012, Hausser_2013}, to perform simultaneous $g$, $r$, and $i$-band fits for each source classified as an LSBG.
PSF information from the parent image tiles was used to generate a Gaussian PSF model for each source.
To identify nucleated systems, we also carried out both single-component S\'ersic fits and combined S\'ersic+PSF fits.

Using the $r$-band images, common masks for bright neighbouring sources across all bands were generated with the Python implementation of {\tt SExtractor} \citep[SEP;][]{sxtractor, Barbary_2016_sep}. These masks were used for the S\'ersic fitting. 
For the purpose of local sky estimation, we applied an additional circular mask centred on the target, with radius equal to the \texttt{FLUX\_RADIUS}. This radius was capped at $5\arcsec$, which (for a $40\arcsec\times40\arcsec$ cutout) limits the masked area to at most $\sim20$ per cent and prevents excessive loss of sky pixels. The combined masks were then used with \texttt{SEP} to estimate and subtract the local sky background. 

Initial parameter guesses were taken from the multi-band catalogue: {\tt MAG\_AUTO} and {\tt FLUX\_RADIUS} for magnitude and half-light radius, {\tt B\_IMAGE}/{\tt A\_IMAGE} for axis ratio (constrained to $0.3 < q \leq 1.0$), and {\tt THETA\_J2000} for position angle (PA). The S\'ersic index ($n$) was initialised at 1 and allowed to vary within $0.2 < n < 5.0$. The centre was fixed at {\tt X=100} and {\tt Y=100} pixels, corresponding to the centre of the cutouts, with an allowed variation of $\pm 10$ pixels ($\sim 2\arcsec$). For the PSF component, magnitudes were also initialised with {\tt MAG\_AUTO} and centres tied to the same coordinates with the same allowed offsets. During the fit, $X$, $Y$, $n$, $q$, PA and $x$, $y$ positions of PSF were tied across all three bands, while magnitudes, sizes, and PSF magnitudes were allowed to vary independently.

 The mean surface brightness (for $g,r$ and $i$ bands) was computed following Eq.~\ref{msb_relation}, while the central surface brightness (for $g,r$ and $i$ bands) was calculated using the relation \citep{Graham_sersic}:
\begin{equation}
\mu_{0} = \overline{\mu}_{\mathrm{eff}} + 2.5\log_{10}\bigg(\frac{n}{b^{2n}}\Gamma(2n)\bigg),
\end{equation}
where $b$ is defined by $2\gamma(2n,b) = \Gamma(2n)$, with $\Gamma$ and $\gamma$ denoting the complete and incomplete gamma functions, respectively. The fitted parameters from {\tt galfitm} were corrected for Galactic extinction and photometric zero-point offsets using the {\tt EXTINCTION\_x} and {\tt D\_MAG\_x} columns from the multi-band catalogue where $x \in \{g, r, i\}$. In addition, we re-applied the colour cuts described in Sect.~\ref{catalog_preselection} with the S\'ersic colours to remove the potential high-$z$ contaminants that have passed through. 
Sources with poor or failed fits (reduced $\chi^{2} > 3$ or unconverged $n$, $q$, $X$, or $Y$ values at parameter limits) were discarded. In addition, to identify duplicates, sources were internally cross-matched using an increasing radius until the number of matches saturated, which occurred at a radius of 5 arcsec. Among the duplicate sources, the one with the lowest reduced $\chi^{2}$ was kept as the final source.

The above-mentioned fitting procedure was performed for both the single S\'ersic and the combined S\'ersic+PSF models, resulting in two sets of structural parameters. To identify nucleated sources based on the S\'ersic+PSF fits, we applied three criteria.
First, the centroids of the S\'ersic and PSF components must lie within $1\arcsec$ of each other, since star-forming regions or unrelated foreground/background sources can otherwise be incorrectly fitted as a PSF component. Second, we require the PSF magnitudes to be brighter than the KiDS~DR5 limiting depths reported by \citet{Wright_2025_kids_dr5}: 
$g = 24.96$, $r = 24.79$, and $i = 23.49$~mag (5$\sigma$, point-source limits).
Third, we assessed the quality of the fits using the reduced $\chi^{2}$ values and the Bayesian Information Criterion (BIC), defined as
\begin{equation}
\mathrm{BIC} = \chi^{2} + n_{\mathrm{param}} \ln(N),
\end{equation}
where $n_{\mathrm{param}}$ is the number of free parameters in the model and $N = n_{\mathrm{dof}} + n_{\mathrm{param}}$ represents the total number of pixels included in the fit, with $n_{\mathrm{dof}}$ denoting the degrees of freedom.
The model with the lower BIC value is considered the better fit. A source is classified as nucleated only if both the reduced $\chi^{2}$ and the BIC of the S\'ersic+PSF model are lower than those of the single S\'ersic fit. 

However, identifying nuclear components in LSBGs using ground-based imaging and S\'ersic+PSF model fitting can also introduce contaminations from spiral galaxies with bulges. Hence, during visual inspection (see Sect.\ref{vis_inspect}), galaxies that are classified as spiral and identified as nucleated candidates were reclassified as non-nucleated in the final catalogue.  
Since the main focus of this study is not the identification of nucleated LSBGs, the catalogue is not refined further, and the reported nucleated sources should be considered only as candidates for nucleated LSBGs. 
After applying the above selections, we obtained 22\,716 LSBG candidates from the \texttt{galfitm} analysis, which were subsequently passed on for visual inspection.

\subsection{Visual inspection}\label{vis_inspect}
We used visual inspection as the final step to improve the purity of the sample. For upcoming large-scale surveys such as LSST, however, visual inspection will not be feasible due to the sheer data volume. In such cases, alternative approaches such as accepting a low-contamination sample ($\sim5\%$) or employing crowd-sourced classifications may be required.
In this work, we selected all the sources that were classified as LSBGs by the models and passed the selection criteria of LSBGs with updated {\tt galfitm} parameters. The resulting candidate sample was then visually inspected by a team of 18 co-authors. To ensure redundancy and consistency, the sample was divided into overlapping subsets such that each source was independently examined by at least three inspectors.

The visual inspection was carried out using a Python-based graphical interface that we developed for this project, which is made publicly available\footnote{\url{https://github.com/lisieckik/LSBMorph}}. The tool allowed the inspectors to display KiDS $r$-band cutouts, colour images made from the KiDS $g$, $r$, and $i$ bands at multiple zoom levels, and the corresponding $r$-band S\'ersic model fits from \texttt{galfitm}. Each candidate was classified into one of three categories: LSBG, non-LSBG (contaminant), or misfitted LSBG. In addition, inspectors were asked to assign a morphological tag following \citet{Lazar_dwarf_morphology}, choosing among four groups: spiral galaxy, elliptical, featureless and confusing or uncertain if the visual inspector is not confident about the morphology.
The final classification label was assigned only when at least two of the three inspectors agreed. For sources without a majority (i.e., when all three labels differed), the 18 co-authors conducted a joint round of visual inspection to determine the final label. In total, 22\,264 LSBG candidates passed the visual inspection.

\subsection{Redshift cross-match with spectroscopic and \ion{H}{i} surveys}\label{redshift_cross_macth}

The catalogue of sources that passed the visual inspection was cross-matched with the spectroscopic catalogues in the KiDS field, such as the 2dF Galaxy Redshift Survey (2dFGRS; \citealt{2dfgres_Colless_2001}), GLADE \citep{Glade_2018}, the Galaxy And Mass Assembly (GAMA; \citealt{Driver_2022_gama}) survey, the SDSS data release 17 \citep{sdss_17_Abdurrouf_2022}, and the Dark Energy Spectroscopic Instrument (DESI; \citealt{DESI_DR1}) survey, using a matching radius of $1\arcsec$. In addition, we cross-matched with \ion{H}{i} surveys such as the Arecibo Legacy Fast ALFA (ALFALFA; \citealt{alfalfa_Haynes_2018}) survey and the Five-hundred-meter Aperture Spherical Radio Telescope (FAST; \citealt{fast_Zhang2024}) survey using a radius of $30\arcsec$. This relatively large radius accounts for the coarse angular resolution of the \ion{H}{i} data (ALFALFA beam $3.8\farcm \times3\farcm3$; FAST beam $2\farcm9$), ensuring counterparts are not missed.

Cross-matches were restricted to spectroscopic and \ion{H}{i} counterparts with redshifts $z \leq 0.2$. This threshold was motivated by clustering analyses showing that LSBGs identified in wide-field surveys such as DES predominantly lie at low redshift ($z \lesssim 0.2$; \citealt{Chicoine_2024}). Hence, matches at higher redshifts are likely to be background systems rather than physical counterparts of our LSBG candidates.
The spectroscopic catalogues used in this work provide survey-specific quality flags that characterise the reliability of redshift measurements. To ensure robust associations, we retained only sources classified as high-confidence redshift determinations according to these indicators. Specifically, we required:
2dFGRS objects to satisfy ${\tt q_z} \ge 3$,
SDSS objects ${\tt ZWARNING}=0$,
DESI objects ${\tt ZCAT\_PRIMARY}>0$ and ${\tt ZWARN}=0$, 
and GAMA objects ${\tt NQ} \ge 3$.

The KiDS field also covers several galaxy clusters. Hence, we cross-matched our LSBG sample with the Meta-Catalogue of X-ray detected Clusters of galaxies (MCXC-II; \citealt{Sadibekova_2024}). All LSBGs located within the angular distance corresponding to the cluster’s $R_{200}$ radius\footnote{We used the $R_{500}$ values and redshifts from \citet{Sadibekova_2024} to compute $R_{200}$, assuming $R_{200} \approx R_{500}/0.65$ following \citet{Ettori2009}. Here, $R_{200}$ is the radius at which the mean density of a cluster is 200 times the critical density of the Universe at that redshift.} were associated with the corresponding cluster. Clusters with redshifts beyond $z = 0.2$ were excluded from the cross-match. For LSBGs that were initially associated with a cluster and had a spectroscopic cross-matched redshift, we further verified their membership in three dimensions. If the distance estimated from the spectroscopic redshift of an LSBG placed it at a radial separation from the cluster centre greater than $R_{200}$, the object was no longer considered a cluster member and was tagged as a non-cluster member.

It is important to note that not all sources located within a cluster’s $R_{200}$ are necessarily physical members; some may be foreground or background galaxies projected along the line of sight. Nevertheless, the majority of these sources are expected to be true cluster members. Therefore, this cluster-associated LSBG sample is best suited for statistical analyses, rather than for studies of individual galaxy properties.

\subsection{Spectral energy distribution fitting}

To model and fit the spectral energy distribution (SED) of observed LSBGs, we used Code Investigating GALaxy Emission \citep[CIGALE, version 2025.0;][]{Boquien20}. The primary goal of the SED fitting is to derive stellar masses ($M_\star$) and star-formation rates (SFRs). CIGALE uses the energy balance conservation between the dust-absorbed stellar emission and its re-emission in the infrared (IR), and it is optimised for broad-band photometry. It goes over a grid of models, with parameter values set by the user, and identifies the best fit by minimising $\chi^2$.
This is followed by the Bayesian approach to determine the optimal properties. Whenever we use CIGALE-derived properties, we refer to the Bayesian value.

The LSBGs in our catalogue have been observed in 9 bands ($u, g, r, i, Z, Y, J, H,$ and $K$). 
For this reason, the flexibility offered by CIGALE is crucial.
The list of the used parameters is presented in Table~\ref{TAB:CIGALE_input}.
We perform two runs using the same parameters: 1) for all LSBGs in our catalogue to estimate the photo-$z$; 2) for the LSBG$_z$ sample (objects with spectroscopic/\ion{H}{1} redshift information, see Table~\ref{tab:number_of_sources}). 
This allows us to check the robustness of the photo-$z$  estimations.
In addition to redshift, we use SED fitting to estimate the stellar masses (M$_\star$) and star formation rates (SFRs, we use SFR averaged over 10 Myr).
We limit the parameters used in the SED fitting to the minimum, focusing only on the stellar population and dust attenuation. This procedure also removes the possibility of SED overfitting. 
We set the star formation history (SFH) to a simple Delayed model with a single burst.
The stellar population has an initial mass function proposed by \cite{chabrier03} with a set subsolar metallicity of 0.08, similar to that reported in previous studies of \ion{H}{II} regions in LSBGs \citep{McGaugh94, deBlok98}.
For dust attenuation, we use the model proposed by \cite{CF2000}.

\section{Results}\label{results}
\subsection{Performance of ML models and validation with known LSBs}

On the labelled dataset of LSBGs and artefacts from DES, the individual DL models achieved a recall of $\sim 95\%$. When combining the model probabilities, a source was classified as an LSBG if at least one model assigned a probability of $P \geq 0.5$. Under this criterion, the ensemble achieved a recall of 97\% with a corresponding contamination rate of $\sim6\%$. By contrast, requiring the average predicted probability to be greater than 0.5 results in a slightly lower recall of 96\%, but a reduced contamination rate of 4\%. Here, we adopt the OR ensemble strategy, as the modest increase in contamination is acceptable given that remaining false positives can be efficiently removed through subsequent visual inspection. 

The receiver operating characteristic (ROC) curves for all models, including the ensemble, are shown in Fig.~\ref{fig:roc}. The precision, recall, accuracy, and area under the ROC (AUROC) curve are listed in Table~\ref{tab:metrics}. Although the models differ in architecture, they achieve similar performance metrics, as shown in Table~\ref{tab:metrics}. The prediction probabilities from each model on the test sample are shown in Fig.~\ref{fig:P_des_dist}. This demonstrates that the models are confident in their predictions, with most outputs close to 0 or 1.

\begin{table}
\caption{Performance metrics of individual models and ensembles at threshold = 0.5.}
\label{tab:metrics}
\centering
\small
\setlength{\tabcolsep}{1pt}
\begin{tabular}{@{}lccccc@{}}
\hline\hline
Model & Precision (\%) & Recall (\%) & Accuracy (\%) & AUROC (\%) \\
\hline
LSBG CNN         & 95.0 & 95.3 & 95.2 & 99.0 \\
LSBG DETR        & 96.3 & 94.5 & 95.5 & 99.2 \\
LSBG ViT         & 95.1 & 95.3 & 95.2 & 99.0 \\
OR Ensemble      & 93.6 & 97.1 & 95.2 & 99.1 \\
Average Ensemble & 96.0 & 95.8 & 95.9 & 99.2 \\
\hline
\end{tabular}
\end{table}

These values indicate the completeness of the reported sample and the level of contamination that must subsequently be mitigated through visual inspection. Since KiDS images differ from DES, the metrics obtained from DES training may not directly reflect performance on KiDS. For example, instrumental artefacts and background characteristics in KiDS images differ substantially from those in DES, which may affect the performance of the DL models. Therefore, the reported contamination rate should be considered as a lower limit for contamination. However, most instrumental contaminations usually result in a poor S\'ersic profile fit and can be filtered out after the S\'ersic profile fitting or subsequent visual inspection. 

Considering that the DES and KiDS surveys overlap over a large area of the sky, we can use the known LSBGs within the KiDS footprint to assess the completeness of our detection. Of the 2,229 known LSBGs in DES within the KiDS footprint, 1,679 (75.3\% of known LSBGs) satisfy the pre-selection criteria described in Sect.~\ref{catalog_preselection}, and 1,596 of these (95.1\% of LSBGs passing the pre-selection) are successfully re-identified as LSBGs in KiDS. Similarly, of the 697 UDGs from \citet{Zaritsky_SMUDGE} within the KiDS footprint, 384 (55.1\% of UDGs) pass the same pre-selection, of which 354 (92.2\% of UDGs passing the pre-selection) are correctly re-identified.

These numbers show that most incompleteness arises during the pre-selection described in Sect.~\ref{catalog_preselection} rather than the ML classification. For example, although the DL ensemble correctly identifies 2\,124 of the 2\,229 DES LSBGs and 561 of the 697 UDGs from \citet{Zaritsky_SMUDGE} when these systems are present in the input sample, many of them never enter the classification stage because they fail the pre-selection cuts (e.g. \re{} < 2\arcsec in the parent catalogue, even though their intrinsic angular sizes are larger than this threshold). Consequently, the final sample is dominated by the limitations of the initial detection catalogue rather than the performance of the ML models.

Unreliable catalogue measurements primarily drive this shortfall at low-surface-brightness regimes. The most cautious problem is shredding, whereby a single diffuse galaxy is incorrectly deblended into multiple smaller components by source-detection algorithms. Such fragmentation leads to biased or unreliable size and surface-brightness estimates, causing these systems to fail the adopted pre-selection criteria. In addition, \citet{Urbano_2025_euclid_completness} demonstrate that catalogue completeness declines sharply at surface brightness levels of $\mu \gtrsim 24~\mathrm{mag \,arcsec^{-2}}$ in the Euclid VIS band. Together, these effects lead to the completeness of our final sample being limited by a combination of surface-brightness–dependent catalogue incompleteness and measurement-related failures in the parent catalogue.

\subsection{Search for LSBGs}\label{Sec:searchForLSBs}
To avoid confusion among the different subsets produced at each stage of the selection process, we explicitly name each subsample defined in this work.
We adopted the KiDS DR5 multi-band catalogue, which contains approximately $1.38 \times 10^{8}$ sources, as the parent sample for identifying LSBGs. After applying the selection criteria on \re{}, \mue{}, $q$, and the $g-r$ and $g-i$ colours, as described in Sect.~\ref{catalog_preselection}, we obtained 322\,278 potential LSBG candidates, hereafter referred to as the pre-selected LSBGs. These sources were then passed to the DL models to estimate their probability of being LSBGs. The resulting probability distributions are shown in Fig.~\ref{fig:P_des_kids}. A source is classified as a candidate if at least one model assigns a probability greater than 0.5. This results in a sample of 28,971 candidates, which we refer to as the deep-learned LSBGs. Examples of sources rejected by the DL models are shown in Fig.~\ref{fig:rejected_sources}.

All sources selected by the models were fitted with both a single-component S\'ersic profile and a combined S\'ersic+PSF model using {\tt galfitm}, as described in Sect.~\ref{sersic_fitting}. After the fitting, we applied the $r$-band cuts, \mue{} $> 23$~\magperarcsec{} and \re{} $> 2.5\arcsec$, together with the axis-ratio and colour cuts (using S\'ersic colours) from Sect.~\ref{catalog_preselection}. In addition, we removed the duplicate sources by an internal cross-match with a $5\arcsec$ radius.  This yielded 22\,716 potential LSBG candidates, which we refer to as S\'ersic-selected LSBGs. These sources are then visually inspected, resulting in a final sample of 22\,264 confirmed objects, hereafter referred to as the visually confirmed LSBGs. 

Sources that passed visual inspection were cross-matched for redshifts as described in Sect.~\ref{redshift_cross_macth}, resulting in 5\,896 sources with spectroscopic redshifts and 1\,102 LSBGs associated with 66 clusters. Among these, 1,006 LSBGs had multiple spectroscopic measurements from different surveys. The cluster-associated redshifts are primarily photometric, with only one cluster-associated LSBG having a spectroscopic redshift. In total, 6\,997 unique sources were found to have redshift estimates, which we call LSBGs with redshift. After correcting for cosmological surface-brightness dimming, we identified 4\,913 LSBGs with redshifts, hereafter referred to as the LSBG$_z$ sample, including 434 UDGs.

\begin{figure} 
    \centering
    \includegraphics[width=\linewidth]{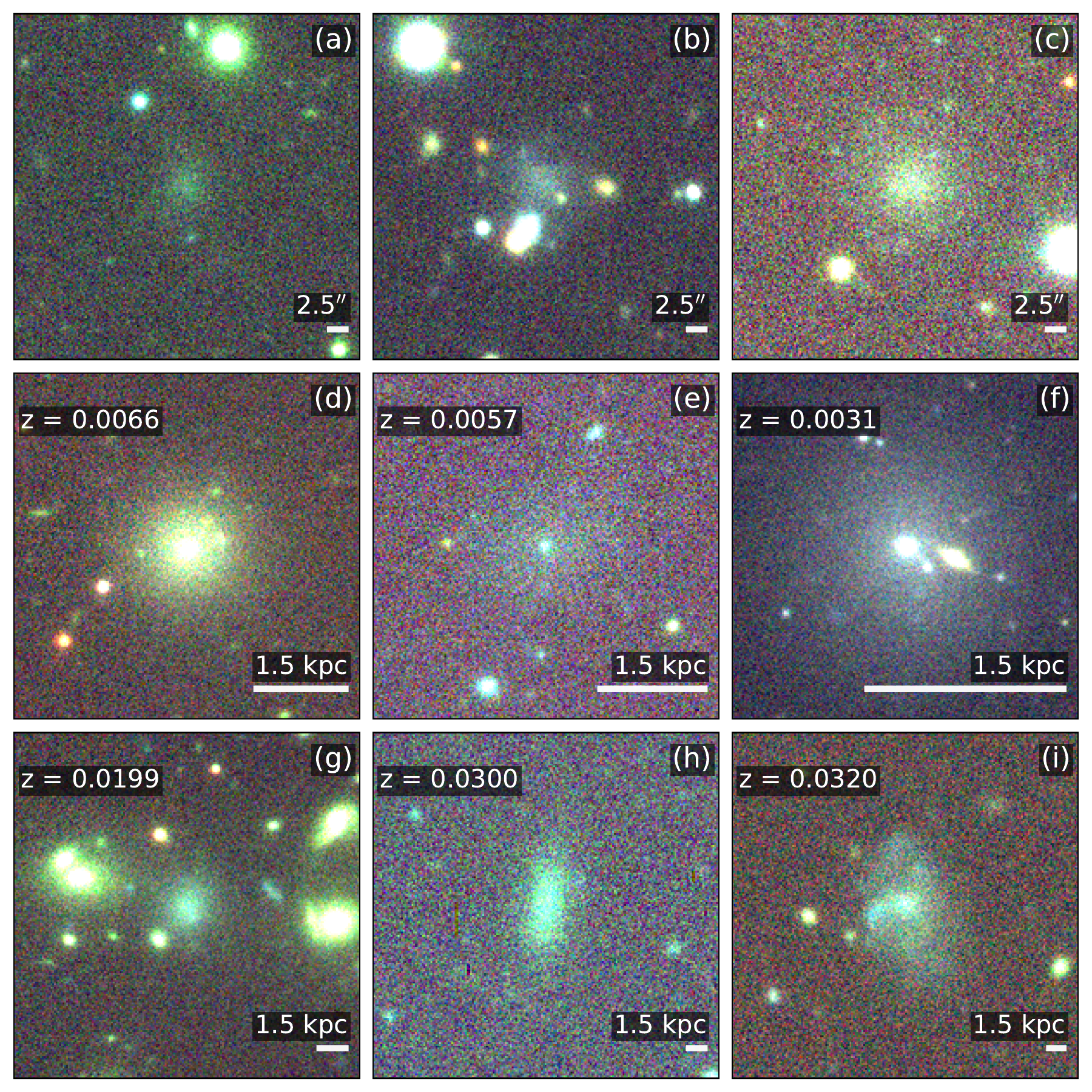}
\caption{
Example RGB images of LSBGs and UDGs created using the $i$, $r$, and $g$ bands with the {\tt APLpy} package \citep{Robitaille}.
Each panel shows a $40\arcsec \times 40\arcsec$ region centred on the source.
Panels (a)--(c) display LSBGs without redshift measurements, panels (d)--(f) show LSBGs with spectroscopic redshifts, and panels (g)--(i) present UDGs with spectroscopic redshifts.
A scale bar indicating either $2.5\arcsec$ (top row) or $1.5$~kpc (middle and bottom rows) is shown in each panel.
}

    \label{fig:kids_lsb_udgs}
\end{figure}

Sources that were reclassified as non-LSBGs after applying the cosmological dimming correction (2\,084 in total) are referred to as the Non-LSBGs$_z$ sample. This sample is provided as a separate supplementary catalogue, enabling future studies to exclude these objects. After excluding these sources from the final catalogue, we report a final sample of 20\,180 LSBGs in KiDS DR5, which is referred to as KiDS-LSBGs. Some examples of LSBGs and UDGs identified in this work are shown in Fig.~\ref{fig:kids_lsb_udgs}. In addition, for all identified sources, photo-$z$ values were estimated along with derived physical parameters such as stellar mass and SFR. When spectroscopic or cluster redshifts were available, these measurements were adopted to compute the above-mentioned physical properties included in the final KiDS DR5 LSBG catalogue.

The number of sources in each sample, along with the selection criteria used to define them, is summarised in Table~\ref{tab:number_of_sources}. The description of the catalogues KiDS-LSBGs and non-LSBGs$_z$ is presented in Appendix \ref{lsbs_table} and \ref{non_lsbs_table}, respectively. Due to the fixed cutout size, the resultant LSBG sample might miss very extended nearby LSBGs. For example, some nearby LSBGs can have angular sizes of $\sim 60$–$90\arcsec$, such as ESO 410-G005 \citep{Karachentsev_2000} in the Sculptor Group (distance $\sim$2 Mpc). As a result, only a fraction of their diffuse light distribution may be included in the cutouts. However, such large-angular-size systems are also absent from the DES training set. Consequently, while this may reduce sensitivity to very nearby or extremely extended LSBGs, it is not expected to introduce a systematic bias relative to the training data.

\begin{table*}[t]
    \caption{Number of sources remaining after each selection step in the construction of the KiDS-LSBG catalogue.}
    \centering
    \begin{tabularx}{\textwidth}{l X r}
        \toprule
        Sample & Definition / Selection & Number \\
        \midrule

        KiDS DR5 parent &
        All detections in the KiDS DR5 multi-band catalogue (9-band photometry) &
        $1.38 \times 10^{8}$ \\

        Pre-selected LSBGs &
        Colour, size, axis-ratio, and surface-brightness cuts (Sect.~\ref{catalog_preselection}) &
        322\,278 \\

        Deep-learned LSBGs &
        Selected by at least one model in the DL ensemble &
        28\,971 \\

        S\'ersic-selected LSBGs &
        Satisfying \mue{} $> 23~\mathrm{mag\,arcsec^{-2}}$ , \re{} $> 2.5\arcsec$ axis ratio and colour cuts from S\'ersic fits &
        22\,716 \\

        Visually confirmed LSBGs &
        Passed visual inspection &
        22\,264 \\

        LSBGs with redshift &
        Total LSBGs with any redshift information (spec-$z$ or cluster) &
        6\,997 \\

        LSBG$_z$ sample &
        After cosmological dimming correction &
        4\,913 \\

        UDGs &
        UDG subset of the LSBG$_z$ sample &
        434 \\

        Non-LSBGs$_z$ &
        Reclassified as non-LSBGs after cosmological dimming correction &
        2\,084 \\

        KiDS-LSBGs &
        Visual-confirmed LSBGs excluding Non-LSBGs$_z$ &
        20\,180\\
        \bottomrule
    \end{tabularx}
    \label{tab:number_of_sources}
\end{table*}

\section{Discussion}\label{discs}
\subsection{Cross-survey validation and comparisons with previous LSBG samples}

\begin{figure}
    \centering
        \begin{subfigure}{\linewidth}
        \centering
        \includegraphics[width=\linewidth]{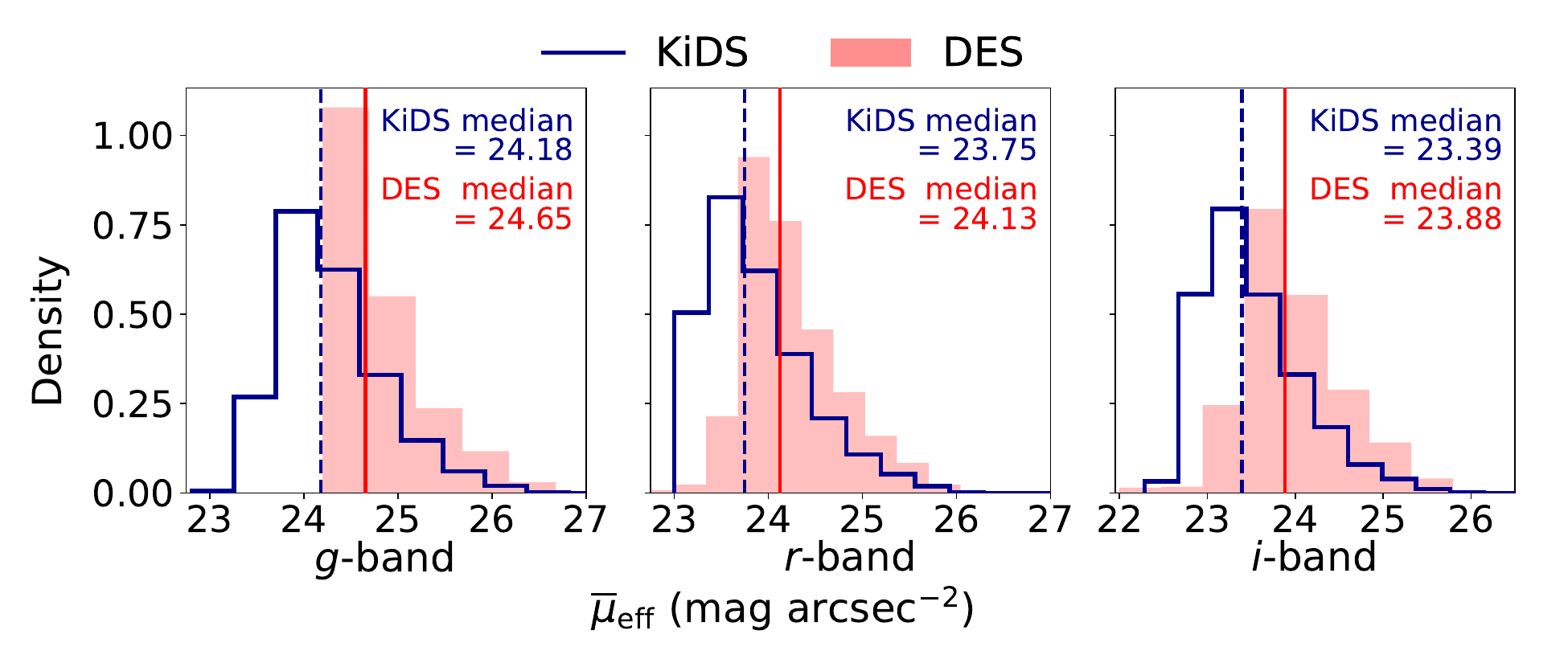}
        \caption{Distribution of $\overline{\mu}_{\mathrm{eff}}$ for the KiDS-LSBGs and DES-LSBGs in the $g$-, $r$-, and $i$-bands.}
        \label{fig:surface_brightness_hist_a}
    \end{subfigure}
    \begin{subfigure}{\linewidth}
        \centering
        \includegraphics[width=\linewidth]{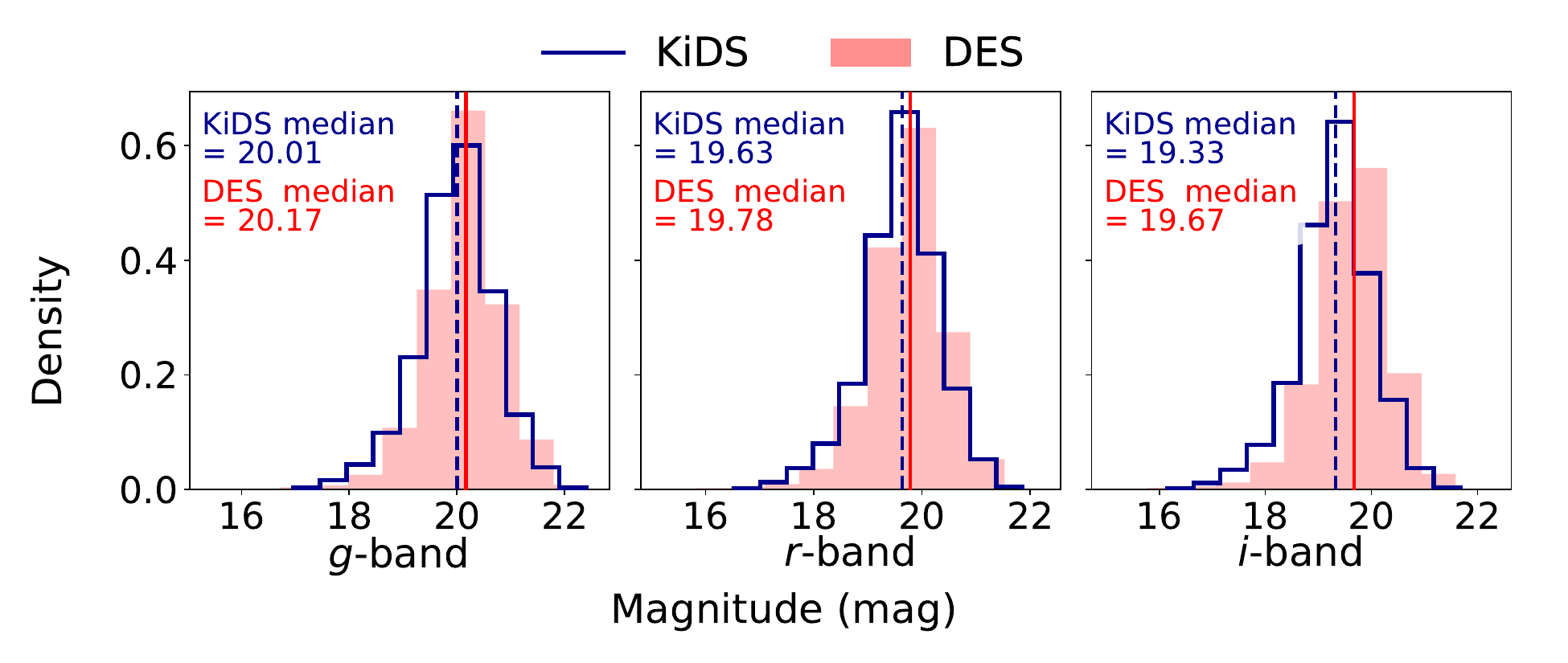}
        \caption{Distribution of magnitudes for the KiDS-LSBGs and DES-LSBGs in the $g$-, $r$-, and $i$-bands.}
        \label{fig:surface_brightness_hist_b}
    \end{subfigure}
    \caption{Each panel shows the normalised density distribution, with KiDS-LSBGs shown in blue and DES-LSBGs in orange. The median surface value for each sample is indicated in the legend of each panel. }
    
    \label{fig:surface_brightness_hist}
\end{figure}

The mean surface brightness distribution of KiDS–LSBGs, compared to that of LSBGs reported in DES (hereafter DES–LSBGs; \citealt{Tanoglidis1, haressh_lsbs}), shows that KiDS–LSBGs are, on average, brighter by approximately $\sim0.5$~\magperarcsec{}, as seen in Fig.~\ref{fig:surface_brightness_hist_a}. Only a small fraction ($\lesssim10\%$) of DES–LSBGs are brighter than the median $r$-band surface brightness of the KiDS–LSBG sample. It should also be noted that DES–LSBGs are defined based on the $g$-band rather than the $r$-band, which may introduce a systematic offset.

However, despite these differences in surface-brightness distributions and selection definitions, the trained DL models successfully identify diffuse systems in this slightly brighter regime. This can be attributed to the overlap between KiDS–LSBGs and DES–LSBGs in other morphological and photometric properties.
For instance, the magnitude distributions of KiDS–LSBGs and DES–LSBGs show significant overlap in the $g$-, $r$-, and $i$-bands, as shown in Fig.~\ref{fig:surface_brightness_hist_b}. The overlap coefficients are $\sim0.89$, $\sim0.89$, and $\sim0.78$ in the $g$-, $r$-, and $i$-bands, respectively. This indicates that both samples occupy a similar nearby region in magnitude space, although the DES–LSBGs are systematically fainter by $0.15$–$0.35$ mag.

Similarly, the structural parameters of LSBGs from KiDS, DES, and HSC (hereafter HSC–LSBGs; \citealt{Greco}), such as $n$, $q$, and \re{}, exhibit consistent distributions and comparable median values within the scatter (Table~\ref{tab:median_structural_params}). While these parameters provide only a simplified description of galaxy morphology and implicitly assume smooth, symmetric light profiles, the observed consistency suggests that the models capture the dominant structural characteristics of LSBGs.

These results indicate that the trained models rely on a combination of features rather than any single feature or definition. As long as there is sufficient overlap in the multi-dimensional feature space between the training and target datasets, the models can generalise well. This allows them to perform reliably even in slightly different regimes, such as in the brighter LSB regime. However, they would struggle to extrapolate when multiple features shift simultaneously beyond the training distribution.

\begin{table} 
  \centering
  \caption{Median structural parameters of KiDS–LSBGs, DES–LSBGs, and HSC–LSBGs. Uncertainties are MAD.}
  \label{tab:median_structural_params}
  \begin{tabular}{lccc}
    \hline
     & $n$ & $q$ & $r_{\mathrm{eff,g}}$ [arcsec] \\
    \hline
    KiDS–LSBGs & $0.95 \pm 0.18$ & $0.71 \pm 0.12$ & $3.48 \pm 0.53$ \\
    DES–LSBGs  & $1.07 \pm 0.25$ & $0.74 \pm 0.11$ & $3.54 \pm 0.60$ \\
    HSC–LSBGs  & $0.86 \pm 0.22$ & $0.70 \pm 0.11$ & $4.08 \pm 0.77$ \\
    \hline
  \end{tabular}
\end{table}

In the framework of the $\Lambda$CDM cosmology, LSBGs are expected to be naturally abundant \citep{Martin_2019, Cintio_2019}. 
Recent surveys have reported varying LSBG number densities depending on selection criteria. In the HSC-SSP, \citet{Greco} found a surface density of $\sim3.9$ deg$^{-2}$ for galaxies with $\overline{\mu}_{\mathrm{eff},g}>24.3$ mag arcsec$^{-2}$, while \citet{Tanoglidis1} and \citet{haressh_lsbs} measured a higher density of $\sim5.5$ deg$^{-2}$ in DES for $\overline{\mu}_{\mathrm{eff},g}>24.2$ mag arcsec$^{-2}$. Both studies adopted an effective radius threshold of $r_{\mathrm{eff},g}>2.5 \arcsec$. 
Applying the same definitions, our KiDS sample yields a surface density of $\sim6.2\,\mathrm{deg}^{-2}$ when using the \citet{Greco} criterion, increasing to $\sim7.3\,\mathrm{deg}^{-2}$ for the \citet{Tanoglidis1} and \citet{haressh_lsbs} selections. This difference reflects our use of a more relaxed {\tt SExtractor}-based preselection size cut ($>2.0\arcsec$; Sect.~\ref{catalog_preselection}), motivated by the $\sim20\%$ underestimation of radii in {\tt SExtractor} measurements \citep{hareesh_2025_hsc,Su_2025_kiDS_udgs}. Adopting a $2.5\arcsec$ {\tt SExtractor} size threshold in the preselection would remove approximately $50\%$ of the KiDS-LSBGs.
These results indicate that current LSBG samples are incomplete, and our quoted surface densities should be regarded as lower limits. Future deep imaging surveys such as Euclid and the LSST are expected to reveal substantially larger LSBG populations.

Recently, \citet{Su_2025_kiDS_udgs} employed a DL model to identify UDGs in KiDS data. Owing to the lack of redshift information, they selected objects with a growth-curve circular effective radius\footnote{The growth-curve circular effective radius in \citet{Su_2025_kiDS_udgs} is defined as the radius enclosing 50\% of the total flux, measured using concentric circular apertures.} in the range $3\arcsec$–$20\arcsec$ and a mean surface brightness \mue{} $> 23.8~$\magperarcsec{} (a conservative threshold) as UDG candidates. Applying these criteria, they identified 545 UDG candidates after excluding spiral morphologies through visual inspection. 
Among these 545 UDGs, only 454 sources were included in the pre-selected LSBGs and entered in our detection pipeline. Among them, 31 sources were not detected by the DL ensemble, and 20 sources failed the combined {\tt galfitm} and visual inspection step. Overall, our final KiDS sample recovers $\sim74\%$ of the UDGs reported by \citet{Su_2025_kiDS_udgs}. When examining the recovery rate at each stage of the pipeline, namely preselection (83.3\%), DL detection (93.2\%), and {\tt galfitm} + visual inspection (95.3\%), it becomes clear that the primary bottleneck here is the preselection step described in Sec.~\ref{catalog_preselection}.

Moreover, the preselection presented in Sec.~\ref{catalog_preselection} is based on the KiDS DR5 parent catalogue, which is not optimised for detecting faint, extended sources. As a result, a fraction of genuine LSBGs may remain undetected, reducing the overall completeness of the sample. For instance, \citet{Xu_2026} report that {\tt SExtractor}-based detection catalogues in surveys such as DES, which are not optimised for faint extended sources, achieve a recovery fraction of $\sim80\%$ at \mue{} $\approx 25~$\magperarcsec{}. The recovery fraction further decreases to $\sim50\%$ at \mue{} $\approx 26~$\magperarcsec{} and declines rapidly at fainter surface brightness levels. 

This behaviour is well known in the LSB regime. Even methods specifically designed for faint, diffuse sources, such as those based on {\tt MTO} \citep{Teeninga_2016_MTO}, are subject to similar limitations. For instance, \citet{Venhola_2026} quantified the completeness of an {\tt MTO}-based detection catalogue in KiDS and showed that the detection efficiency depends strongly on both the surface brightness and the size of the source (see Fig.~2 in \citet{Venhola_2026}). In particular, beyond \mue{} $\approx 26~$\magperarcsec{}, the completeness of the detection catalogue typically drops below $\sim50\%$. Since reprocessing the entire KiDS dataset is beyond the scope of this work, we do not attempt to address these limitations here.

However, even with these limitations, our sample is substantially larger than that reported by \citet{Su_2025_kiDS_udgs}. For instance, using the same KiDS DR5 dataset and applying similar cuts on the circularised effective radius\footnote{The circularised effective radius provides a reasonable proxy for the size definition adopted by \citet{Su_2025_kiDS_udgs}.} and \mue{} along with the removal of spiral galaxies, we identify 3,276 UDG candidates. The smaller sample size reported in their study likely reflects methodological differences, including the detection strategy, size measurements, and criteria used during visual inspection. Consequently, the two samples are not directly comparable. Nevertheless, our sample is approximately five times larger, suggesting that our approach may recover additional UDG candidates that were not identified in their search.

Within the KiDS-LSBGs, a small level of contamination from background galaxies is also expected due to cosmological surface brightness dimming. Approximately 10\% of our initial visually inspected sample, which had cross-matched redshifts, were subsequently removed after applying dimming corrections. To estimate the maximum possible impact on the full sample, we conservatively assume a common redshift of $z=0.11$. This corresponds to the 90th percentile of the spectroscopic LSBG redshift distribution and maximises the contribution of higher-redshift contaminants. Under this conservative assumption, we estimate that up to $\sim26\%$ of the sample could be affected, which should be interpreted as a conservative upper limit. Even in this extreme case, our UDG candidate sample would be reduced to $\sim 2400$ sources, which remains more than three times larger than the sample reported by \citet{Su_2025_kiDS_udgs}.

\subsection{KiDS-LSBGs}
\begin{figure}[h]
    \centering
    \includegraphics[width=\linewidth]{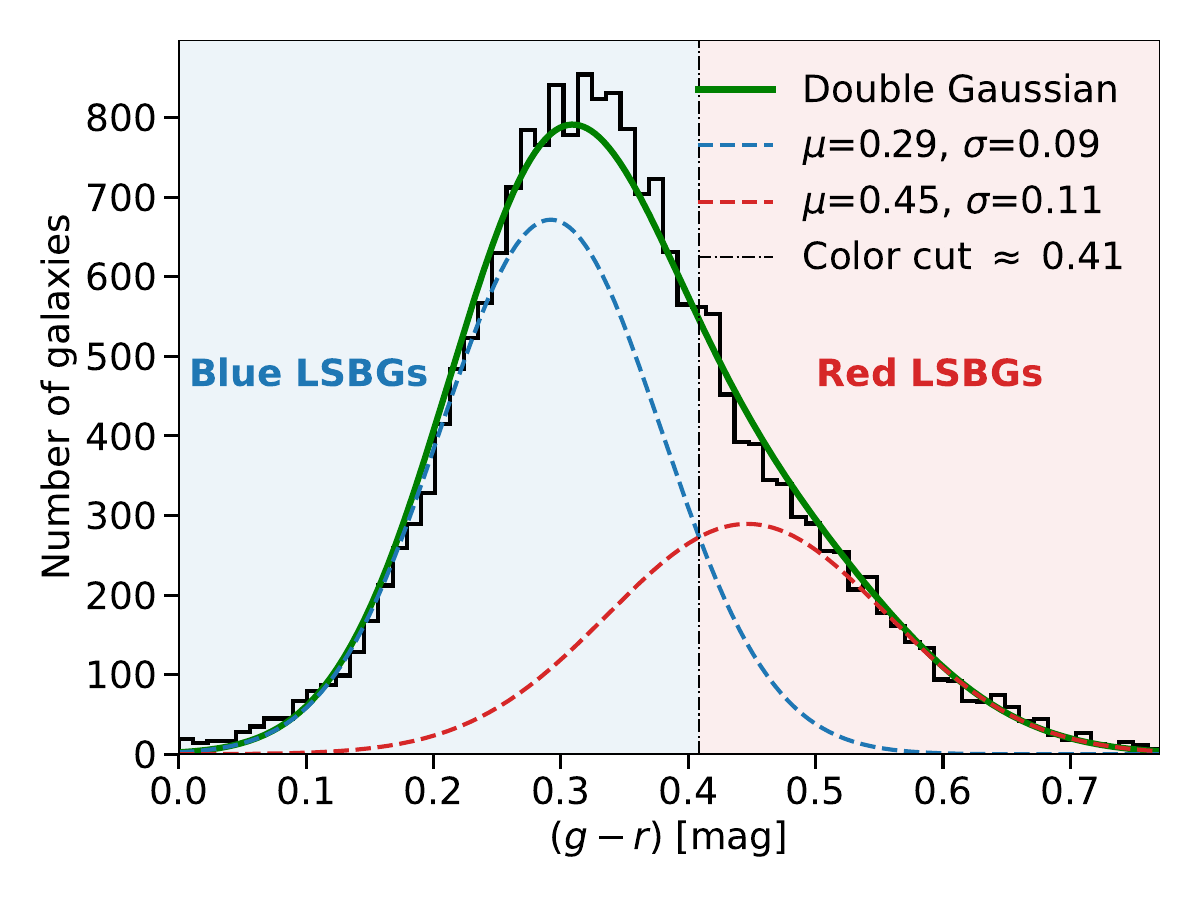}
    \caption{Distribution of $(g-r)$ colours for the KiDS-LSBG sample. The black step histogram shows the colour distribution, while the green curve represents the best-fitting double Gaussian model. The blue and red dashed curves denote the individual Gaussian components, corresponding to the blue and red LSBG populations, respectively. The mean ($\mu$) and the standard deviation ($\sigma$) of each Gaussian component are shown in the legend. The vertical dashed–dotted line indicates the adopted colour cut at $(g-r)=0.41$, which separates the two populations. The shaded regions show the division between blue and red LSBGs.}
    \label{fig:colour_dist}
\end{figure}

The optical $(g-r)$ colour distribution of the KiDS-LSBG sample is shown in Fig.~\ref{fig:colour_dist}. We modelled the distribution using both single and double Gaussian fits and compared the models using the Bayesian Information Criterion (BIC). A lower BIC indicates a better fit, and a difference of $\Delta$BIC $\geq 100$ can be decisive evidence in favour of the model with the smaller value \citep{Kass_BIC}. For our sample, the double-Gaussian model yields a lower BIC than the single-Gaussian case, with a difference of 568, providing very strong evidence for bimodality. Based on this fit, we adopt $(g-r)=0.41$ as the dividing colour between blue and red LSBG populations. An analogous fit to the $(g-i)$ colour distribution yields a division at $(g-i)=0.74$, with the double-Gaussian model strongly favoured over a single-Gaussian fit ($\Delta\mathrm{BIC}=563$). Applying these colour cuts independently, we find that the KiDS sample comprises approximately $73\%$ blue and $27\%$ red LSBGs in both cases, consistent with previous studies \citep{Greco, Tanoglidis1, Du_2024}.

\begin{figure}
    \centering
    \includegraphics[width=\linewidth]{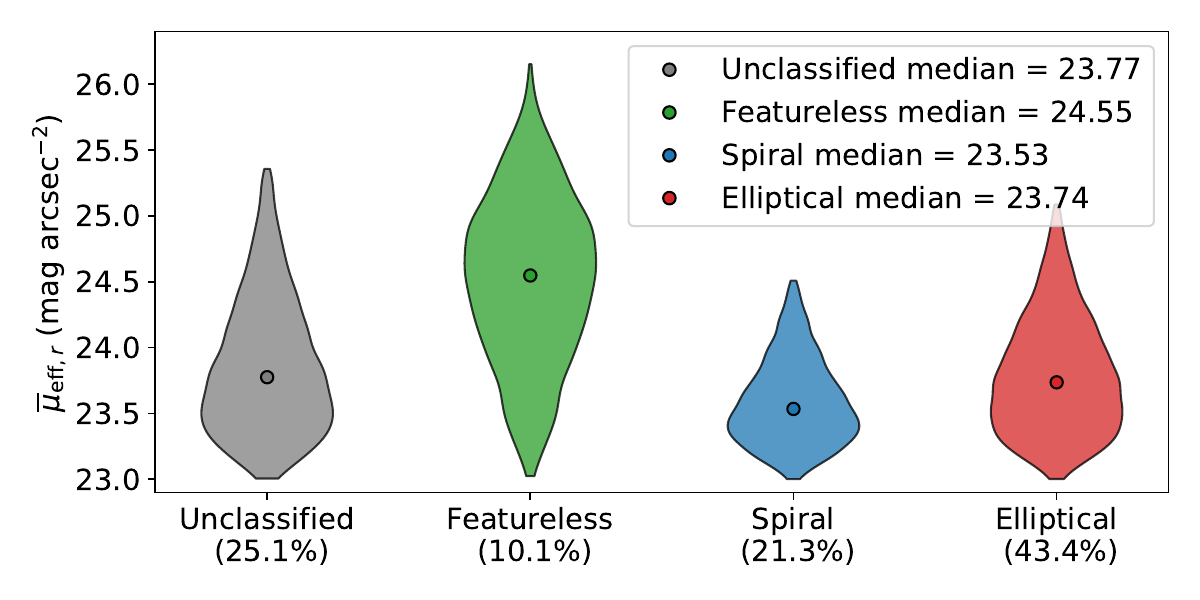}
    \caption{Mean effective $r$-band surface brightness as a function of visual morphology. The sample is divided into unclassified (25.1\,\%), featureless (10.1\,\%), spiral (21.3\,\%), and elliptical (43.4\,\%) galaxies. The horizontal lines indicate the median surface brightness for each class.}
    \label{fig:Morphology_surface_brightness}
\end{figure}

Having established the colour bimodality of the sample, we next examine how the structural properties of LSBGs vary with morphology. The mean $r$-band surface brightness as a function of morphology is shown in Fig.~\ref{fig:Morphology_surface_brightness}. During visual inspection, morphological classification was not mandatory; consequently, $25.1\,\%$ of the sources lack a morphology label due to ambiguity, lack of consensus, or irregular appearance. Of the classified galaxies, $43.4\,\%$ are ellipticals, $21.3\,\%$ are spirals, and $10.1\,\%$ are featureless systems. Spiral and elliptical galaxies exhibit similar surface-brightness distributions, with median \mue{} values of $23.53$ and $23.74~\mathrm{mag\,arcsec^{-2}}$, respectively. In contrast, featureless galaxies are, on average, fainter, with a median \mue{} of $24.55~\mathrm{mag\,arcsec^{-2}}$. Likewise, the median S\'ersic indices are comparable for ellipticals ($n = 0.97$) and spirals ($n = 0.99$), while featureless systems exhibit systematically lower values ($n = 0.80$). We note that higher spatial resolution imaging may alter the morphological classifications presented here and could consequently affect the reported median structural parameters. Overall, elliptical and spiral galaxies display broadly similar structural parameter distributions, whereas featureless galaxies differ slightly on average. These trends are consistent with previous studies of dwarf galaxy morphologies (e.g. \citealt{Lazar_dwarf_morphology}). 

However, these colour and morphology classifications are based on ground-based imaging and are therefore affected by PSF limitations. To assess this effect, we examined $155$ galaxies in the overlapping region between the KiDS footprint and Euclid Q1 imaging \citep{Euclid_Q1_Aussel}. Among the $68$ galaxies classified as ellipticals in KiDS, $24$ are found to be spiral galaxies in the higher-resolution Euclid images, exhibiting very faint spiral arms and, in many cases, disturbed morphologies. Accounting for this reclassification, the spiral fraction in the overlap region increases to approximately $45\,\%$, while the elliptical fraction decreases to about $28\,\%$. This comparison indicates that, while the KiDS-based morphological classifications capture the broad structural properties of the galaxies, higher-resolution imaging can reveal additional low-contrast features that are not readily detectable in ground-based data. Consequently, the reported morphology fractions should be interpreted as representative of the dominant visible structures at KiDS resolution. 

Furthermore, initially, about $\sim 10\%$ of the KiDS-LSBG sample was classified as nucleated based on the S\'ersic+PSF modelling described in Sect.~\ref{sersic_fitting}.
A comparison with Euclid Q1 imaging \citep{Euclid_Q1_Aussel} revealed that several of the fitted “nuclei” correspond to bulges of spiral galaxies rather than true nuclear star clusters. Out of 155 KiDS-LSBGs with available Euclid imaging, 24 were fitted as nucleated, but visual inspection confirmed only four as truly nucleated. The remaining sources were mostly spiral galaxies with prominent bulges. In addition, two nucleated sources were classified as non-nucleated, corresponding to a nucleated fraction of about 4\%. This demonstrates a limitation of using single S\'ersic$+$PSF fits for identifying nucleated LSBGs in ground-based data.

\subsection{Size-luminosity relation for LSBGs and UDGs}\label{photo-size_luminosty}
\begin{figure} 
    \centering
    \includegraphics[width=\linewidth]{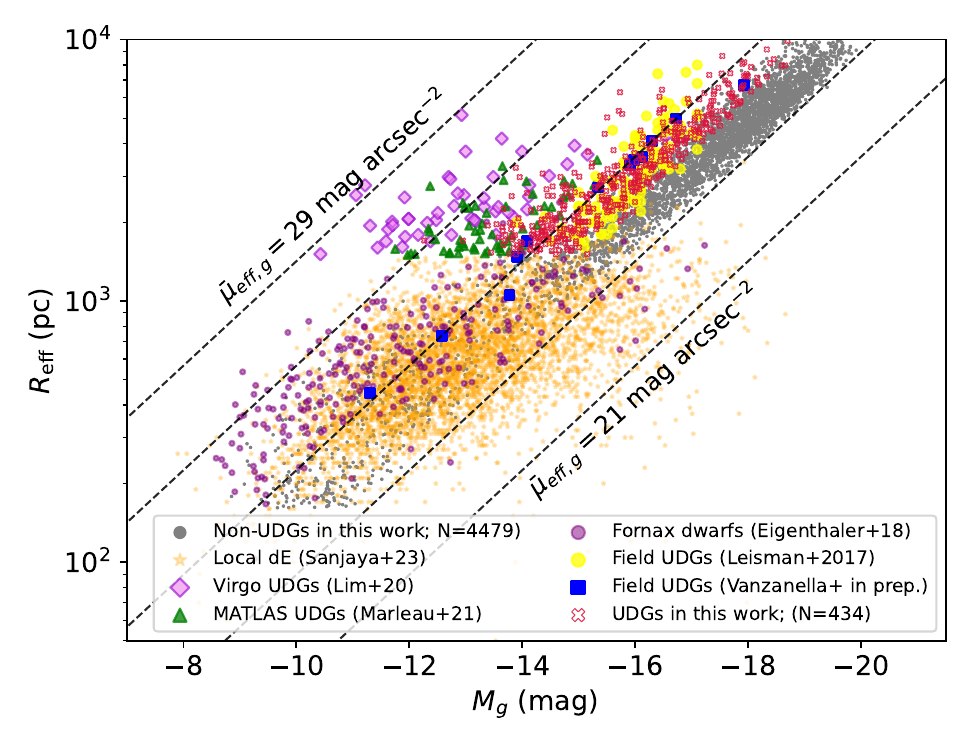}
    \caption{
The size-luminosity relation for the non-UDGs and UDGs compared with dwarf galaxies and UDGs from the literature.  The effective radius ($R_{\rm eff}$) is shown as a function of absolute $g$-band magnitude ($M_g$). Non-UDGs are shown as grey points, while KiDS UDGs are highlighted with red open crosses. Local dwarf ellipticals from \citet{Paudel_sanjaya} and Fornax cluster dwarfs from \citet{Eigenthaler_2018} are shown as orange stars and purple filled circles, respectively. UDGs in the Virgo cluster \citep{Lim} and from the MATLAS survey \citep{Marleau} are indicated by violet diamonds and green triangles, respectively. Field UDGs from \citet{Leisman_2017} and  Vanzanella et al. (in prep.) are shown as yellow-filled circles and blue-filled squares, respectively. Dashed black lines indicate loci of constant mean effective surface brightness in the $g$ band, ranging from $\bar{\mu}_{\rm eff,g} = 21$ to $29~\mathrm{mag~arcsec^{-2}}$, with adjacent lines separated by 2~mag~arcsec$^{-2}$.
}\label{fig:size_luminosty_relation}
\end{figure}

The structural properties, such as the S\'ersic index and axis ratio of dwarf galaxies, are very similar to the values reported in Table \ref{tab:median_structural_params} \citep{Poulain}. Simulations further indicate that the dark matter halos of UDGs and dwarf galaxies have comparable masses \citep{Amorisco, Benavides_2023}. These results suggest that the LSBGs and UDGs found in this work could be an extended population of dwarf galaxies. A comparison of the size-luminosity relations for local dwarf ellipticals \citep{Eigenthaler_2018, Paudel_sanjaya} and UDGs \citep{Lim, Marleau} from the literature, the non-UDGs (objects classified as LSBGs but not satisfying the UDG definition) and UDGs that have a cross-matched redshift from this work, is presented in Fig. \ref{fig:size_luminosty_relation}.

Although the non-UDGs and UDGs identified in this work are, on average, more luminous than local dwarf ellipticals, they remain diffuse due to their systematically larger effective radii. This implies that these systems are more massive than typical local dwarf galaxies, while maintaining comparable mean stellar surface densities. The non-UDGs and UDGs preferentially populate the brighter end of the size–luminosity plane relative to cluster UDG samples from \citet{Lim} and \citet{Marleau}, which is expected given that those samples originate from targeted follow-up observations with deep data.
In contrast, the present sample is drawn from a blind cross-matching of KiDS-LSBGs with existing redshift surveys, naturally biasing the detected population toward brighter and more readily observable systems. As a result, our UDG sample shows closer agreement with field UDGs from \citet{Leisman_2017}, which were also identified through a blind H I-selected survey.

Overall, non-UDGs (LSBGs), UDGs, and local dwarf galaxies define a continuous structural sequence in the size–luminosity plane. This continuity suggests that LSBGs and UDGs do not constitute a distinct galaxy class, but rather represent diffuse extensions of the dwarf-galaxy population, shaped by a combination of internal processes and environmental effects \citep{Conselice, Marleau, Wang_2023, Bautista, zoller_coma}.

\subsection{SED fitting of LSBGs}\label{sed_fitting_results}
\subsubsection{Photo-z estimation}\label{photo-z_estiamtion}
Dedicated spectroscopic surveys are typically used to measure the redshifts of galaxies observed in large-scale photometric surveys. However, because of their faint and diffuse nature, LSBGs and UDGs are particularly difficult targets for spectroscopy \citep{Du_2017, Roman_2021}. For example, nearly 85\% of the spectroscopically cross-matched LSBGs in our sample have $\bar{\mu}_{eff,r} < 24~\mathrm{mag~arcsec^{-2}}$. This highlights that, even with extensive spectroscopic programs such as DESI \citep{DESI}, only a small fraction of the faintest LSBG and UDG populations can be probed. While dedicated spectroscopic follow-up campaigns could, in principle, address this limitation, such observations would be extremely time-consuming and impractical on the scale required for surveys such as LSST or Euclid.

In this context, photo-$z$ provide a natural alternative. These redshifts can be estimated using broad-band photometry with either empirical (typically machine-learning–based methods; \citealt{Bilicki_2018, Anjitha_2025}) or SED fitting techniques \citep{Bolzonella_2000}. In this work, we use SED fitting with CIGALE to estimate photo-$z$ values for visually confirmed LSBGs, employing nine photometric bands available from KiDS ($u$, $g$, $r$, $i$) and VIKING ($Z$, $Y$, $J$, $H$, $K_s$). The LSBG$_z$ sample was used to assess the reliability of the estimated photo-$z$ values.

\begin{figure}
    \centering
    \includegraphics[width=\linewidth]{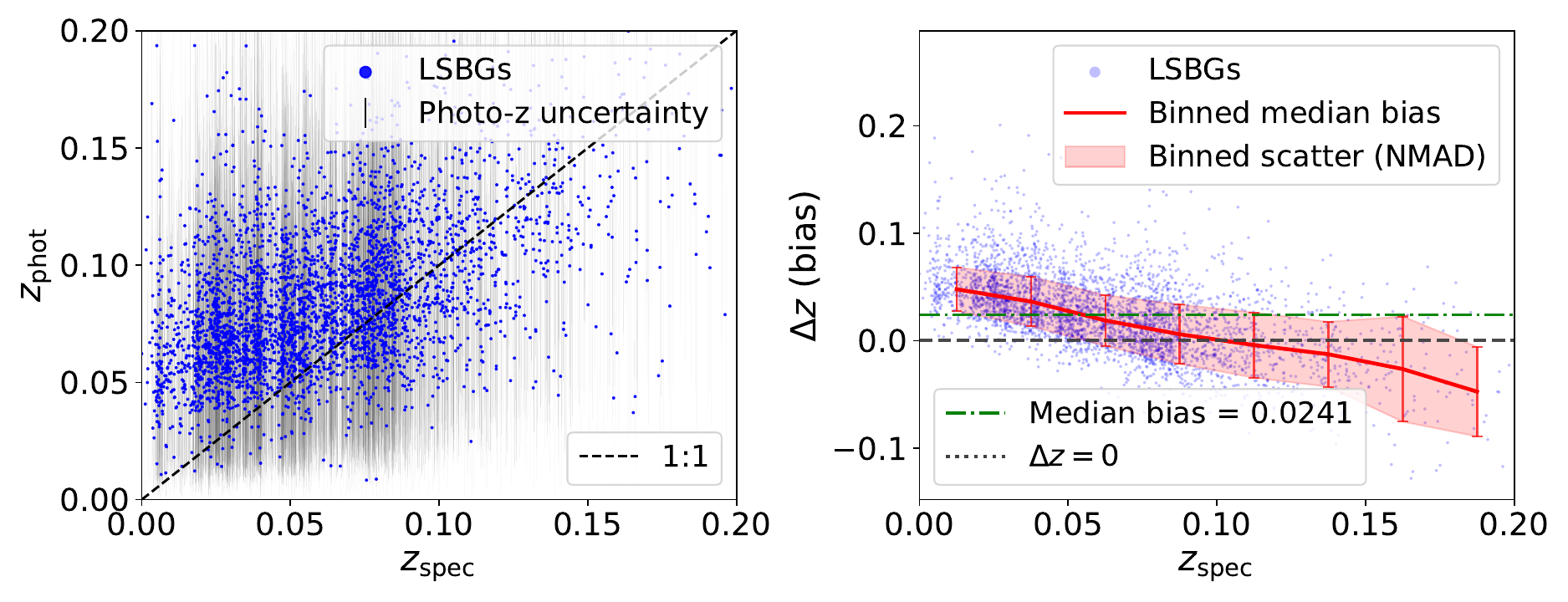}
    \caption{The estimated photo-$z$ as a function of spectroscopic redshift is shown in the left panel. The dashed line indicates the one-to-one relation, while the vertical error bars represent the photo-$z$ uncertainties. The right panel shows the redshift bias, defined as $\Delta z = z_\mathrm{phot} - z_\mathrm{spec}$, as a function of spectroscopic redshift. The red curve denotes the binned median bias, and the shaded region represents the scatter quantified via the NMAD within each bin. The green dash-dotted line marks the median bias of the full sample, while the black dotted line indicates $\Delta z = 0$.}
    \label{fig:photo_z}
\end{figure}

The estimated photo-$z$ ($z_\mathrm{phot}$) as a function of spectroscopic redshift ($z_\mathrm{spec}$), along with the corresponding redshift bias ($\Delta z = z_\mathrm{phot} - z_\mathrm{spec}$) as a function of $z_\mathrm{spec}$, are shown in Fig.~\ref{fig:photo_z}.
On average, the estimated photo-$z$ values overestimate the redshifts, with $\Delta z\simeq 0.024$. The bias exhibits a mild redshift dependence, with photo-$z$ values tending to be overestimated at the lowest redshifts ($0 < z < 0.1$) and progressively underestimated toward higher redshifts ($z > 0.1$). Such behaviour is commonly observed in photo-$z$ analyses \citep{Bilicki_2018, Bilicki_2021, Anjitha_2025} and is expected given the non-uniform redshift distribution of the reference LSBG population \citep{Francis_Peacock_2010}.

The photo-$z$ estimates also show relatively large uncertainties for individual sources. This behaviour is expected since LSBGs and UDGs are predominantly nearby systems, and photo-$z$ methods generally perform less reliably at very low redshifts \citep{Bordoloi_2010, Hildebrandt_2012}. At low redshift, small errors in colour measurements can translate into comparatively large fractional redshift uncertainties. Moreover, LSBGs are characterised by diffuse light distributions and low signal-to-noise ratios, which further degrade photometric accuracy.

The scatter in the photo-$z$ estimates is quantified using the normalised median absolute deviation (NMAD), defined as:
\begin{equation}
\sigma_\mathrm{NMAD} = 1.48 \times \mathrm{median} \left( \frac{|\Delta z - \mathrm{median}(\Delta z)|}{1 + z_\mathrm{spec}} \right),
\end{equation}
where $\Delta z = z_\mathrm{phot} - z_\mathrm{spec}$. 
The NMAD value obtained for our sample ($\sigma_\mathrm{NMAD} = 0.031$) is comparable to that reported for KiDS-DR4 photo-$z$ estimates in \citet{Bilicki_2018}. In that work, $\sigma_{\mathrm{NMAD}} = 0.025$ for the redshift bin $0 < z < 0.1$ and $\sigma_{\mathrm{NMAD}} = 0.034$ for $0.1 < z < 0.2$. For the same redshift ranges, our photo-$z$ estimates yield $\sigma_{\mathrm{NMAD}} = 0.027$ and $\sigma_{\mathrm{NMAD}} = 0.028$, respectively. This agreement indicates that, despite the challenging photometric properties of LSBGs, the overall photo-$z$ performance remains broadly consistent with that achieved for the general KiDS galaxy population.

Nevertheless, future surveys such as LSST and Euclid will require further refinement of photo-$z$ estimation techniques for LSBGs and UDGs, whose diffuse nature presents persistent challenges for accurate redshift determination. In particular, improvements in photometric depth, wavelength coverage, and modelling of LSBG SEDs will be crucial for mitigating systematic biases and reducing redshift uncertainties in these systems.

\subsubsection{Main Sequence of LSBGs}\label{main_sequence_subsection}
Generally, the global star formation rate (SFR) of star-forming galaxies correlates with their stellar mass \citep{Brinchmann_2004, Noeske_2007, Speagle_2014, Popesso23}. This relation, known as the star-forming main sequence (MS), is commonly expressed as
\begin{equation}\label{main_sequence}
    \log(\mathrm{SFR}) = a\,\left[\log(M_\star)-8.5\right] + b,
\end{equation}
where $M_\star$ is the stellar mass, $a$ and $b$ are free parameters. 
Here, $10^{8.5}$ M$_\odot$ is chosen as the pivot mass for normalisation, as it better reflects the characteristic mass scale of our LSBG sample. Selecting a pivot mass near the centre of the sampled mass distribution reduces covariance between the fitted parameters and improves the stability of the regression \citep[e.g.][]{Speagle_2014, Pearson_2018, Pearson_2023}. The position of a galaxy relative to the MS provides insight into its evolutionary state, as departures from the sequence can be used to identify systems that are quenching or undergoing enhanced star formation \citep[e.g.][]{Elbaz18, Donevski20, Lisiecki25}.

\begin{figure}  \centering \includegraphics[width=\linewidth]{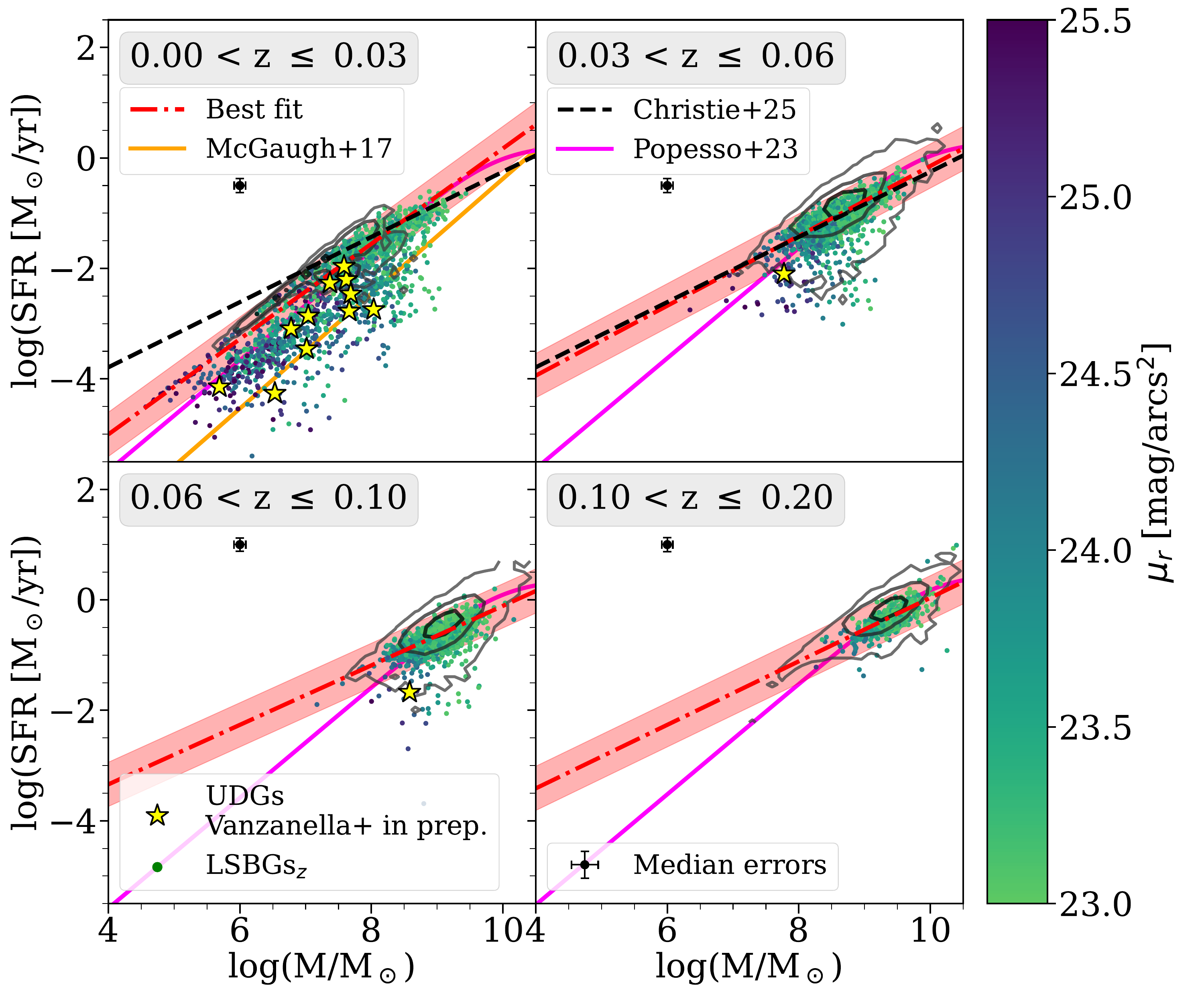} \caption{SFR as a function of $M_\star$ for LSBGs. Each panel corresponds to a different redshift bin, indicated in the upper-left corner. Coloured points represent the LSBG$_z$ sample. The median error within the LSBG$_z$ sample is shown as a black point. Grey contours show the distribution of LSBGs with photo-$z$, marking the 95th and 99th percentiles. The red dash-dotted line with red shaded region ($\pm 0.4$ dex) illustrates the best linear fit for LSBGs with spectroscopic redshifts. Yellow stars mark the positions of field UDGs from  Vanzanella et al. (in prep.). For comparison, known LSBG relations from the literature are shown: orange line \citep{McGaugh17}, black dashed line \citep{Christie25} and Magenta line \citep{Popesso23}.} \label{fig:MS_LSBs} \end{figure}

To characterise the MS for LSBGs (hereafter referred to as MS$_{\mathrm{LSBGs}}$), we use the LSBG$_z$ subsample with specific star formation rates (sSFR), satisfying $\log\,\mathrm{sSFR}\,[\mathrm{yr}^{-1}] > -11$ \citep[the threshold originates from][]{salim18}. The relation is then fit using Eq.~\ref{main_sequence} in each redshift bin. Excluding low-sSFR galaxies ensures that the fit is constrained by genuinely star-forming systems. The resulting parameters are reported in Table~\ref{tab:MSfit}.
In Fig.~\ref{fig:MS_LSBs} we present the SFR-$M_\star$ relation for the LSBGs identified in this work, derived from SED fitting and shown in four redshift bins. Each panel includes both the spectroscopic subsample (LSBG$_z$) and the photo-$z$ LSBG sample. 
The colour scale indicates that, at fixed $M_\star$, LSBGs with fainter surface brightnesses (larger \mue{}) systematically occupy the lower envelope of the MS. This trend arises because fainter, more diffuse galaxies have shallower gravitational potentials and less centrally concentrated gas, which leads to lower star formation efficiencies.

\begin{table}
  \centering
  \caption{MS$_{\text{LSBGs}}$ fit parameters for each redshift bin.}
  \label{tab:MSfit}
  \begin{tabular}{ccccc}
    \hline
    $z$ range & $a$ & $b$ & N$_{zphot}$ & N$_{z}$ \\
    \hline
    0.00 -- 0.03 & 0.86 $\pm$ 0.01 & -1.12 $\pm$ 0.01 & \phantom{1}\,666 &1\,611\\
    0.03 -- 0.06 & 0.63 $\pm$ 0.01 & -1.10 $\pm$ 0.01 & 3\,711 &1\,328\\
    0.06 -- 0.10 & 0.54 $\pm$ 0.01 & -0.92 $\pm$ 0.01 & 6\,782 &1\,403\\
    0.10 -- 0.20 & 0.58 $\pm$ 0.02 & -0.84 $\pm$ 0.02 & 4\,609 &\phantom{0}\,571\\
    \hline
  \end{tabular}
  \tablefoot{The first column shows the redshift range. This is followed by two columns referring to parametrisation in Equation~\ref{main_sequence}. The last two columns inform about the number of sources in photo-$z$ and spec/\ion{H}{I}-$z$ sample, respectively.}
\end{table}

A comparison of MS$_{\mathrm{LSBGs}}$ with the dwarf-galaxy MS defined by \citet{McGaugh17} reveals that, at fixed $M_\star$, our LSBGs exhibit systematically higher SFRs.  This difference is likely due to the use of different SFR tracers. \citet{McGaugh17} derived SFRs from H$\alpha$ luminosities, whereas our estimates are based on broadband SED fitting. A similar offset between different SFR indicators was also reported by \citet{Christie25}. This discrepancy may be related to bursty star formation histories and will be investigated in future work.

Comparing the star formation rates predicted by the  MS relation of \citet{Christie25} with the MS${_\mathrm{LSBGs}}$ for galaxies in the lowest redshift bin yields an RMS difference of 0.37 dex. This offset is largely driven by differences in the slopes ($|\Delta a| \simeq 0.27$), reflecting the distinct stellar-mass ranges used to derive the relations. The sample in \citet{Christie25} primarily includes galaxies with $M_\star \gtrsim 10^8,\mathrm{M_\odot}$, whereas our sample extends to much lower masses ($M_\star \sim 10^6,\mathrm{M_\odot}$). Restricting the comparison to $10^8 \leq M_\star \leq 10^{9},\mathrm{M_\odot}$ reduces the RMS difference to $\sim$0.08 dex. Hence, the discrepancy arises from extrapolating the MS of \citet{Christie25} beyond its calibrated range. This is further supported by the $z = 0.00$–0.06 bins, where RMS differences are only 0.04 dex, and our sample overlaps in stellar mass with \citet{Christie25}.

A similar trend is seen when comparing MS${_\mathrm{LSBGs}}$ with the MS of \citet{Popesso23}. In the lowest redshift bin, the RMS difference is 0.23 dex, decreasing to 0.08 dex for $10^8 \leq M\star \leq 10^9,\mathrm{M_\odot}$. At higher redshifts, RMS differences (0.14, 0.11, and 0.09 dex) gradually decline, indicating improved agreement. Overall, discrepancies with literature MS relations are minor within the calibrated stellar-mass ranges. Apparent disagreements mainly result from extrapolation beyond these ranges rather than intrinsic differences in the scaling relations.

While the MS$_{\mathrm{LSBGs}}$ relations derived using spectroscopic redshifts provide the most robust characterisation of the star-forming properties of our sample, the majority of LSBGs lack spectroscopic measurements and therefore we rely on photo-$z$ values. It is therefore important to assess how photo-$z$ uncertainties propagate into derived physical parameters such as stellar mass and SFR.

\begin{figure}
    \centering
    \includegraphics[width=\linewidth]{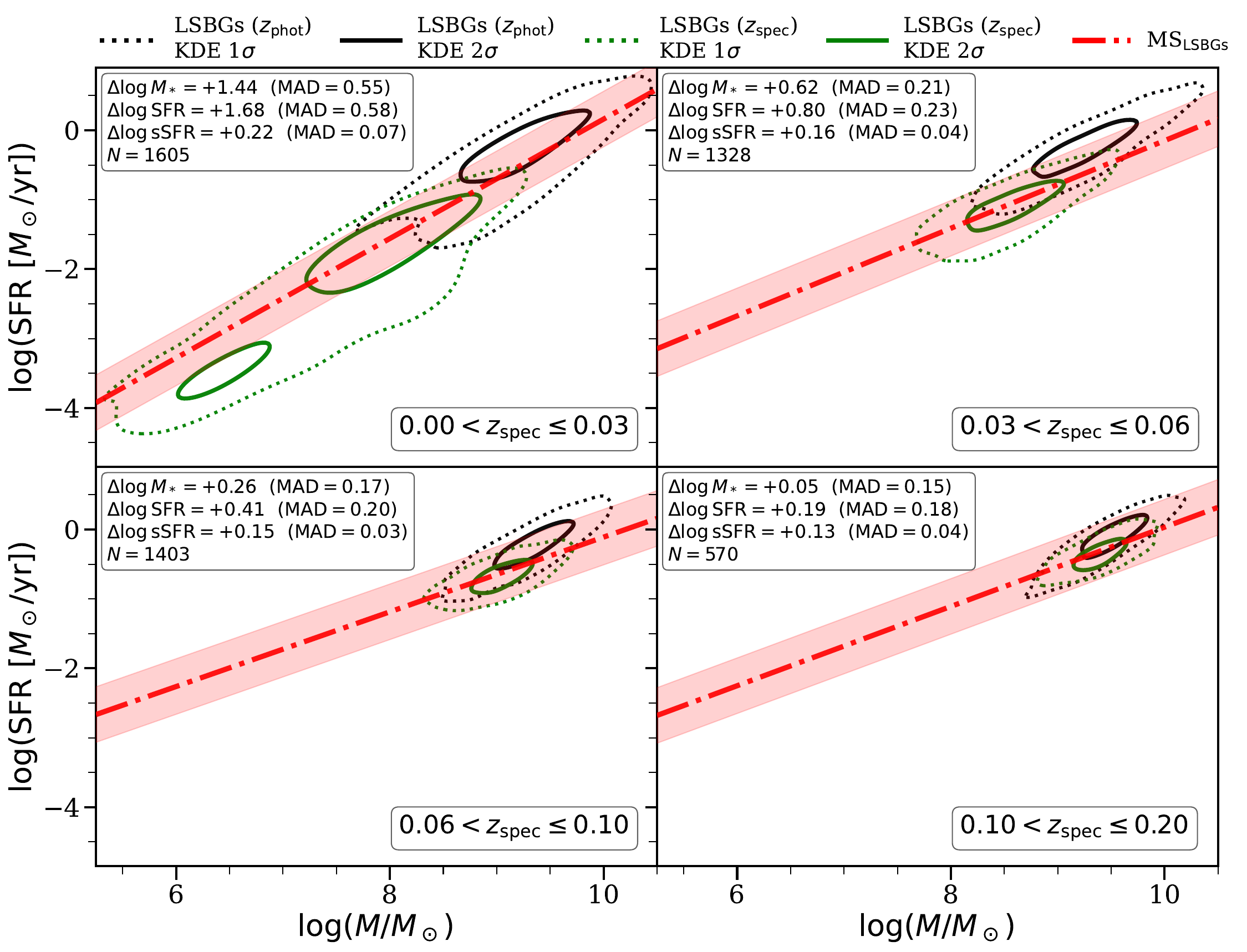}
    \caption{Comparison of the stellar mass and star formation rate (SFR) distributions of LSBGs derived using photo-$z$ values (black contours) and spectroscopic redshifts (green contours). The sample is divided into four photometric-redshift bins. The dotted and solid contours represent the 1$\sigma$ and 2$\sigma$ kernel density estimates, respectively. The median offsets in stellar mass, SFR and sSFR measured in each bin, along with their MAD values, are indicated within the panels. The red dash-dotted line shows the main sequence relation fitted for LSBGs, while the shaded region denotes a $\pm0.4$ dex scatter.}
    \label{fig:MS_photoz_specz}
\end{figure}

The comparison of stellar mass and SFR derived using photometric and spectroscopic redshifts is shown in Fig.~\ref{fig:MS_photoz_specz}. Across all redshift bins, the photo-$z$-based distributions (black contours) are systematically overestimated and offset relative to the spec-$z$-based distributions (green contours). These offsets occur primarily along the star-forming main sequence, as indicated by the relatively small shifts in $\log \mathrm{sSFR}$ ($\sim 0.22$–$0.13$ dex). This behaviour reflects biases in the photo-$z$ estimates: overestimated redshifts lead to inflated luminosity distances and consequently higher inferred luminosities.
The corrections required for the photo-$z$-based stellar masses and SFRs for sources lacking spectroscopic redshifts are presented in Table~\ref{tab:photoz_corrections}. They are derived as the median offsets between photometric and spectroscopic-based measurements. The corrections are largest in the lowest-redshift bin and remain significant across the full redshift range, with substantial scatter. However, despite these biases, the overall structure of the main sequence is preserved.

Hence, photo-$z$ uncertainties do not fundamentally alter the observed star-formation trends but mainly affect the absolute scaling of the derived physical parameters. This is because both stellar mass and SFR scale approximately with the square of the luminosity distance \citep{Kennicutt_1998}. Hence, the biases from photo-$z$ shift the galaxies coherently along the main sequence. However, the impact of photo-$z$ uncertainties on the other relations, such as the size–luminosity relation, is qualitatively different. The absolute magnitude is logarithmically proportional to the luminosity distance, whereas the physical size is linearly proportional to the angular diameter distance \citep{Hogg_1999}. As a result, photo-$z$ errors introduce shifts that are not aligned with the true size–luminosity relation of LSBGs.
These results indicate that, while photo-$z$ values are suitable for studying combined relations such as the main sequence, caution is required when interpreting individual parameters or their distributions, as well as relations involving quantities with different dependencies on distance.

\begin{table}
\centering
\caption{Median correction terms for stellar mass and star formation rate as a function of photometric redshift.}
\label{tab:photoz_corrections}

\setlength{\tabcolsep}{3pt}

\begin{tabular*}{\linewidth}{@{\extracolsep{\fill}}cccc}
\hline
$z_{\mathrm{phot}}$ bins & Corr$_{\log M_\star}$ (dex) & Corr$_{\log \mathrm{SFR}}$ (dex) & $N$ \\
\hline
0.00 -- 0.03 & $-0.41 \pm 0.72$ & $-0.60 \pm 0.81$ & 116 \\
0.03 -- 0.06 & $-0.58 \pm 0.43$ & $-0.75 \pm 0.47$ & 1021 \\
0.06 -- 0.10 & $-0.54 \pm 0.38$ & $-0.72 \pm 0.42$ & 2316 \\
0.10 -- 0.20 & $-0.60 \pm 0.40$ & $-0.79 \pm 0.42$ & 1359 \\
\hline
\end{tabular*}
\tablefoot{The first column lists the photometric redshift range. The second and third columns give the corrections to the stellar mass (Corr$_{\log M_\star}$) and star formation rate (Corr$_{\log \mathrm{SFR}}$), respectively, derived from photometric redshift estimates. The values are reported as the median $\pm$ median absolute deviation (MAD). The offsets are defined as $\Delta \equiv (\mathrm{phot} - \mathrm{spec})$ and represent the systematic corrections to be applied in each $z_{\mathrm{phot}}$ bin. The final column lists the number of sources in each bin.}
\end{table}

It is important to note that our sample is affected by incompleteness and selection biases, particularly at higher redshifts. By construction, the catalogue includes only galaxies with angular sizes larger than $2.5\arcsec$. At $z=0.06$, this angular cut corresponds to a physical size of $\sim2.9$ kpc, which is nearly twice the typical effective radius of a UDG. As a result, increasingly diffuse galaxies fall below our detection threshold at higher redshift. Since neither this work nor \citet{Christie25} explicitly quantifies the completeness of the LSBG sample, the MS$_{\rm LSBGs}$ relation presented here should be regarded as indicative rather than definitive. A more complete census will be required to robustly constrain the MS behaviour of the full LSBG population.

\subsection{Quenched LSBGs and UDGs}
Despite the limitations discussed above, the derived MS$_{\rm LSBGs}$ relation captures several important aspects of LSBG evolution. Most notably, it reveals a substantial population of LSBGs with suppressed star formation ($\Delta{\rm MS}_{\rm LSBGs}<-0.4$ dex) at low redshift. We refer to these systems as “quenched” from here onwards, although they may still retain some residual star formation. In the lowest redshift bin ($z\leq0.03$), quiescent galaxies comprise $\sim42\%$ of the LSBG$_z$ sample, but this fraction rapidly decreases to $\sim5\%$ at $z>0.06$. This decline may be driven by increasing incompleteness, as quiescent LSBGs become progressively more difficult to detect at higher redshift.
Splitting the sample by environment shows that quiescent LSBGs are preferentially found in clusters. At $z\leq0.03$, $72.7\%$ of cluster LSBGs are quenched, compared to $11.1\%$ in non-cluster environments. Although the absolute quenched fractions decrease at higher redshifts, cluster galaxies consistently exhibit higher quiescent fractions than their non-cluster counterparts.

\begin{figure}
    \centering
    \includegraphics[width=\linewidth]{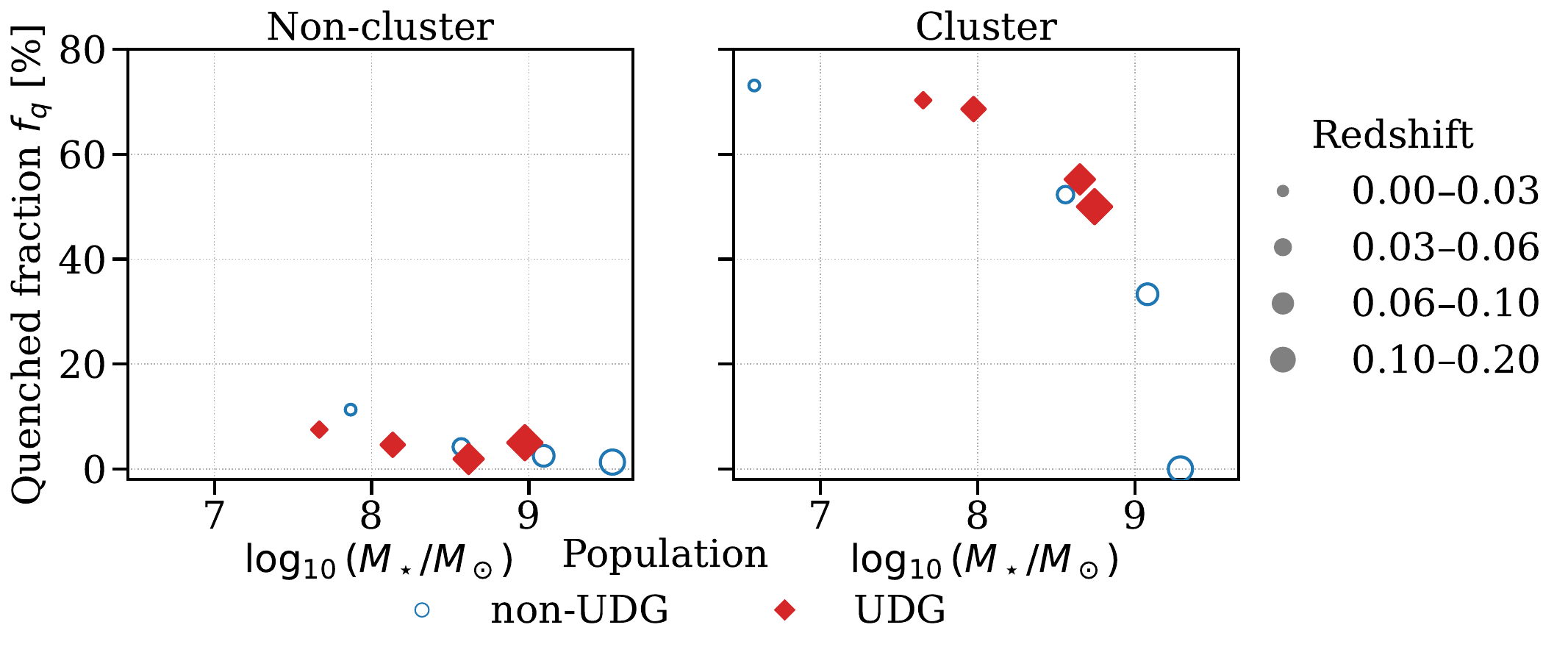}
    \caption{Quenched fraction ($f_q$) versus median stellar mass for non-UDGs (blue open circles) and UDGs (red filled diamonds) in cluster and non-cluster environments. Different marker sizes indicate distinct redshift bins, increasing with redshift from $0.00 < z \leq 0.03$ to $0.10 < z \leq 0.20$. Quenched galaxies are defined by $\Delta{\rm MS}_{\rm LSBGs} < -0.4$ dex.}
    \label{fig:quenched_enviornment}
\end{figure}

The quenched fractions of non-UDGs and UDGs in cluster and non-cluster environments across different redshift bins as a function of the stellar mass are shown in Fig.~\ref{fig:quenched_enviornment}. The corresponding numerical values are listed in Table~\ref{tab:quenched_fractions}.
Due to their more diffuse nature, UDGs are preferentially quenched compared to less diffuse non-UDGs in clusters. In addition, the median stellar masses of UDGs are systematically lower than those of non-UDGs at fixed redshift. Together, these trends indicate that within the LSBG population, environmental quenching is more effective for galaxies with lower stellar mass and lower surface brightness. This environmental dependence is consistent with simulations of LSBGs \citep{Martin_2019} and mirrors trends observed in the general galaxy population, including high-surface-brightness galaxies, where quenching is more efficient at lower stellar masses \citep{Peng_2010}.

\begin{figure} 
    \centering
    \includegraphics[width=\linewidth]{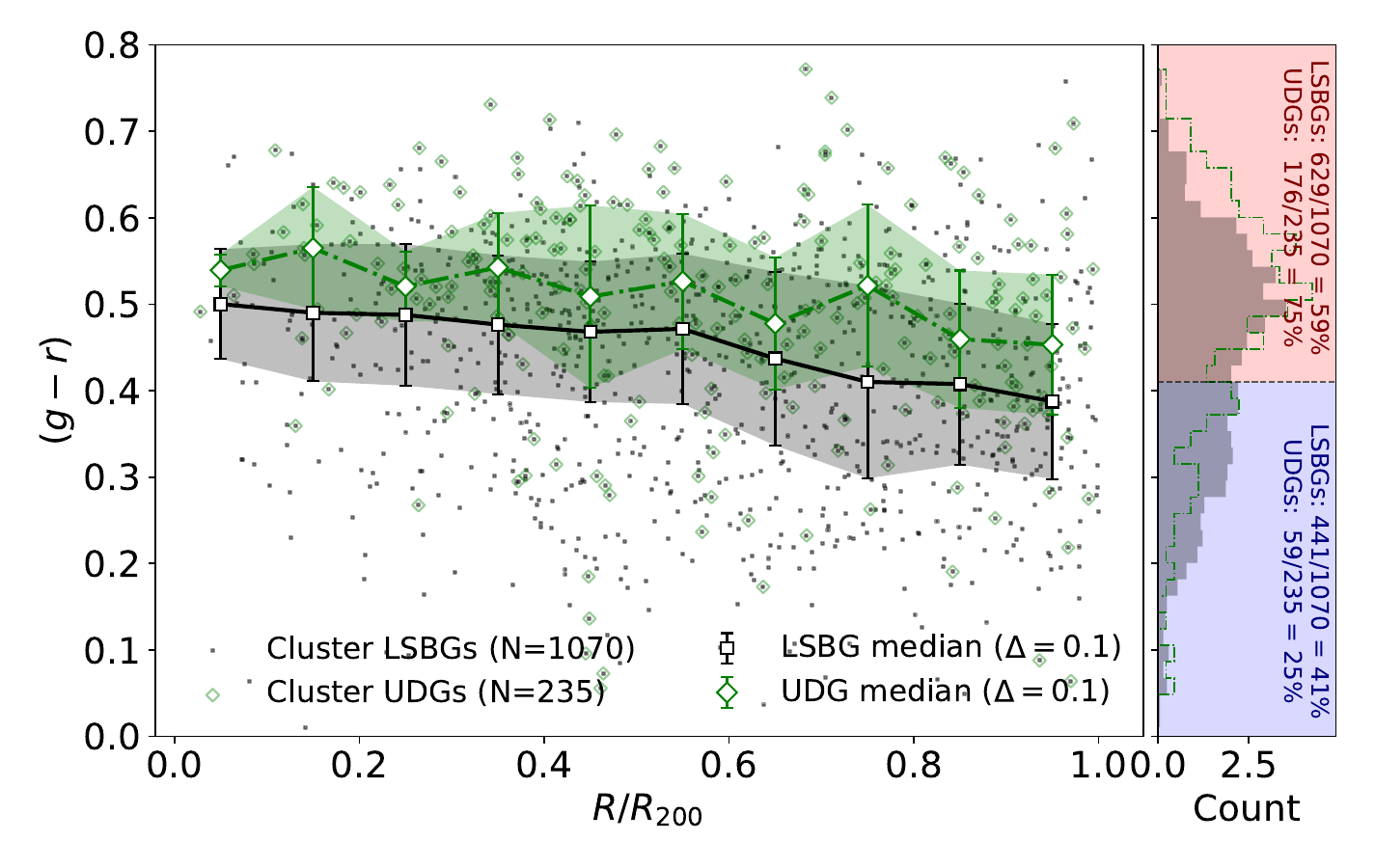}
\caption{The optical $g-r$ colour for LSBGs (black square) and UDGs (green hollow rhombus) found in this work that are associated with clusters as a function of cluster-centric distance (normalised by $R_{200}$). Median values in each radial bin are overplotted, with the shaded regions representing the median absolute deviation. The side panel shows the normalised y-axis histogram of $g−r$ for LSBGs and UDGs. The colour threshold at $g−r=0.41$ partitions the populations into red and blue. The corresponding red/blue fractions for each sample are printed inside the shaded regions.}    \label{fig:g_r_colour_in_cluster}
\end{figure}

A characteristic signature of environmental quenching is the reddening of galaxies toward the cluster centre. This behaviour is evident in the optical $g-r$ colours of the cluster-associated LSBGs and UDGs shown in Fig.~\ref{fig:g_r_colour_in_cluster}. This trend suggests that gas stripping and quenching processes become increasingly effective toward the cluster core, leading to suppressed star formation and redder colours. The large scatter in colour within each radial bin indicates that the impact of the environment varies among clusters, depending on their dynamical state and local conditions. Nonetheless, toward the innermost regions, the smaller uncertainties in the median colours of both populations point to a more uniform environmental influence close to the cluster centre.

In addition, UDGs are consistently redder than LSBGs at all cluster-centric distances, with a significantly higher fraction of red UDGs (75\%) compared to red LSBGs (59\%). This difference implies that UDGs are more strongly affected by environmental quenching, likely due to their lower surface densities and greater susceptibility to tidal and ram-pressure effects. These colour trends are consistent with previous studies of cluster UDGs \citep{Mancera_udg, Junais2022} and support theoretical models in which tidal harassment and ram-pressure stripping play central roles in their evolution \citep{Moore1996, Aguerri}.

Quiescent UDGs and non-UDGs are also detected in non-cluster environments, particularly in the lowest redshift bin, indicating that environmental processes alone cannot account for all quenched systems. These galaxies may be satellites in low-mass haloes or systems that have experienced quenching driven by internal mechanisms such as gas depletion or stellar feedback. This interpretation is supported by simulations of low-mass galaxies \citep{Hayward_2017}, UDG formation models \citep{Chan_2018}. However, the quenched fractions reported here are lower than the $\sim26\%$ quenched fraction of UDGs reported by \citet{Prole}, likely reflecting differences in the adopted quenching criteria, as \citet{Prole} classified all red UDGs as quenched.

The trends and results discussed above rely on the assumption that all sources with a projected distance less than $R_{200}$ are part of the cluster. However, as mentioned earlier in Sec.~\ref{redshift_cross_macth}, some of the LSBGs tagged as cluster members may instead be foreground or background galaxies. For instance, \citet{Xu_2026} show that up to $\sim20\%$ of galaxies identified as cluster or group members (based on a projected distance $< R_{200}$) may be background sources projected along the line of sight (see Appendix~C of \citealt{Xu_2026}). However, given the consistency of the reported trends with the literature, this contamination is expected to primarily affect the strength of the trends rather than significantly alter their overall behaviour. Since obtaining spectroscopic redshifts for all LSBGs in clusters is challenging due to their faint and diffuse nature, our analysis is limited in this regard. Furthermore, as individual photo-$z$ estimates are biased, especially in the local universe, we refrain from refining the cluster membership based on photo-$z$ estimates.

\section{Conclusions}\label{conclusion}

In this work, we demonstrated that with physics-motivated domain adaptation, DL models originally trained on DES imaging can be used to identify LSBGs in the KiDS. After applying surface-brightness and scale normalisation, the models successfully identified over 20\,180 LSBGs and 434 ultra-diffuse galaxies (UDGs) in KiDS DR5. This establishes the robustness and cross-survey nature of this methodology. The recovered structural parameters of the KiDS–LSBGs closely match those obtained from DES and HSC catalogues, confirming that the models capture fundamental morphological properties of diffuse galaxies rather than survey-specific features.

The KiDS–LSBG sample extends the known LSBG population toward higher surface brightnesses, effectively bridging the gap between classical dwarf galaxies and more diffuse systems such as UDGs. In the size–luminosity plane, dwarf galaxies, LSBGs, and UDGs define a continuous structural sequence. This continuity suggests that LSBGs and UDGs do not constitute a distinct galaxy class, but rather represent the diffuse, large-size extension of the dwarf-galaxy population.
 
By cross-matching with spectroscopic catalogues, we obtained reliable redshifts for 4\,913 LSBGs, enabling the systematic characterisation of the star-forming main sequence (MS$_{\text{LSBGs}}$). The derived MS$_{\text{LSBGs}}$ relations are consistent with the main sequence measured by \citet{Popesso23}, as well as with the LSBG main sequence reported by \citet{Christie25}. We note, however, that the choice of star formation rate indicator introduces systematic variations in the inferred main sequence normalisation, consistent with the trends reported by \citet{Christie25}.

For sources lacking spectroscopic redshifts, we estimated photometric redshifts via SED fitting using CIGALE. The achieved photo-$z$ accuracy ($\sigma_\mathrm{NMAD} = 0.031$) is comparable to published KiDS results obtained with dedicated DL photo-$z$ methods. However, the estimated photo-$z$ values are systematically overestimated, which results in inflated stellar mass and star formation rate estimates. Despite this bias, the photo-$z$ solutions remain broadly consistent with the star-forming main sequence. By applying the correction terms presented in Table~\ref{tab:photoz_corrections}, photo-$z$ estimates can therefore be reliably used for population-level analyses, although individual galaxy measurements should be interpreted with caution.

The colour distribution of the KiDS–LSBGs is clearly bimodal ($\Delta\mathrm{BIC}_{g-r} = 563$), with approximately $73\%$ blue and $27\%$ red galaxies. Combined with the observed radial colour gradients in clusters, this behaviour indicates that environmentally driven quenching represents a dominant evolutionary pathway. Cluster LSBGs and UDGs exhibit systematically redder colours and reduced star formation relative to their non-cluster counterparts, consistent with the effects of ram-pressure stripping and tidal interactions. The strong environmental dependence of quenching, particularly for diffuse systems, highlights the importance of jointly considering structural, photometric, and environmental diagnostics when interpreting LSBG evolution.

Overall, this study illustrates the power of domain adaptation for LSBG science. As forthcoming deep surveys, such as Euclid and the Vera C.~Rubin Observatory’s LSST, are expected to reveal millions of diffuse galaxies, survey-specific model training will become increasingly impractical. The framework presented here provides a scalable, homogeneous, and computationally efficient pathway for constructing cross-survey LSBG catalogues and for studying their structural, stellar population, and environmental properties across cosmic time.

\begin{acknowledgements}
    This research was supported by the Polish National Science Centre grant 2023/50/A/ST9/00579. H.T gratefully acknowledges support from the Consolidaci\'on Investigadora IGADLE project with reference CNS2024-154572. He also acknowledges support from the GEELSBE2 project with reference PID2023-150393NB-I00 funded by MCIU/AEI/10.13039/501100011033 and the FSE+, and from the Department of Education, Junta de Castilla y Le\'on and FEDER Funds (Reference: CLU-2023-1-05). J is funded by the European Union (Widening Participation, ExGal-Twin, GA 101158446). K.L. acknowledges the support of the National Science Centre, Poland, through the PRELUDIUM grant UMO-2023/49/N/ST9/00746. K.M. and S.P. acknowledge support from the grant UMO-2024/53/B/ST9/00230 funded by the National Science Centre, Poland. W.J.P., S.D, and A.P.C have been supported by the Polish National Science Center project UMO-2023/51/D/ST9/00147. MF acknowledges support from the SONATA grant of the National Science Center No. 2022/47/D/ST9/00419.
    Based on data obtained from the ESO Science Archive Facility with DOI: https://doi.org/10.18727/archive/37, and https://doi.eso.org/10.18727/archive/59 and on data products produced by the KiDS consortium. The KiDS production team acknowledges support from: Deutsche Forschungsgemeinschaft, ERC, NOVA and NWO-M grants; Target; the University of Padova, and the University Federico II (Naples). We made use of spectroscopic redshift data from DESI, GAMA, 2dFGRS, ALFALFA, and FAST, as well as imaging data from DES DR1 and Euclid Q1. The relevant survey and data release papers are cited in the text.
\end{acknowledgements}
\bibliographystyle{aa}
\bibliography{aa}
\begin{appendix}

\section{Training dataset}
\begin{figure}[h] 
  \centering
  \begin{subfigure}{0.45\linewidth} % About half the width
    \centering
    \includegraphics[width=\linewidth, keepaspectratio]{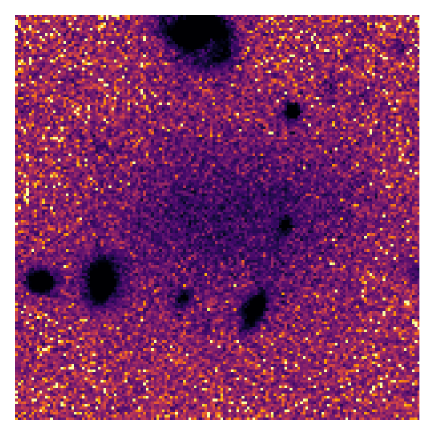}
    \caption{}
    \label{lsb_train}
  \end{subfigure}
  \hfill
  \begin{subfigure}{0.45\linewidth} % About half the width
    \centering
    \includegraphics[width=\linewidth, keepaspectratio]{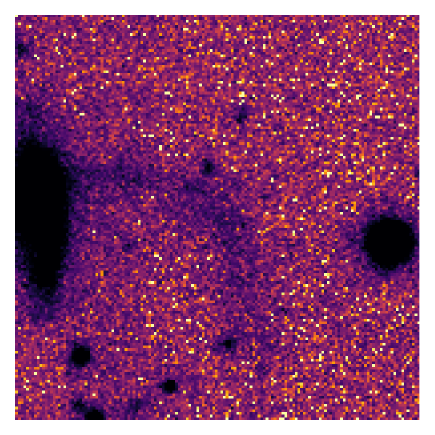}
    \caption{}
    \label{art_train}
  \end{subfigure}
  \caption{\textit{r}-band images of an LSBG (\ref{lsb_train}) and a contaminant (\ref{art_train}) from DES that are used in the training data. Each cut-out corresponds to a $40\arcsec \times 40\arcsec$ ($152 \times 152$ pixels) region of the sky centred on the LSBG or contaminant.}
  \label{fig:training}
\end{figure}

Examples of an LSBG (\ref{lsb_train}) and an artefact (\ref{art_train}) used in the training dataset.

\section{KiDS surface brightness depth}\label{appendix:surface_brightness_depth}
\begin{figure}[h]
    \centering
    \begin{subfigure}{0.95\linewidth}
        \centering
        \includegraphics[width=\linewidth]{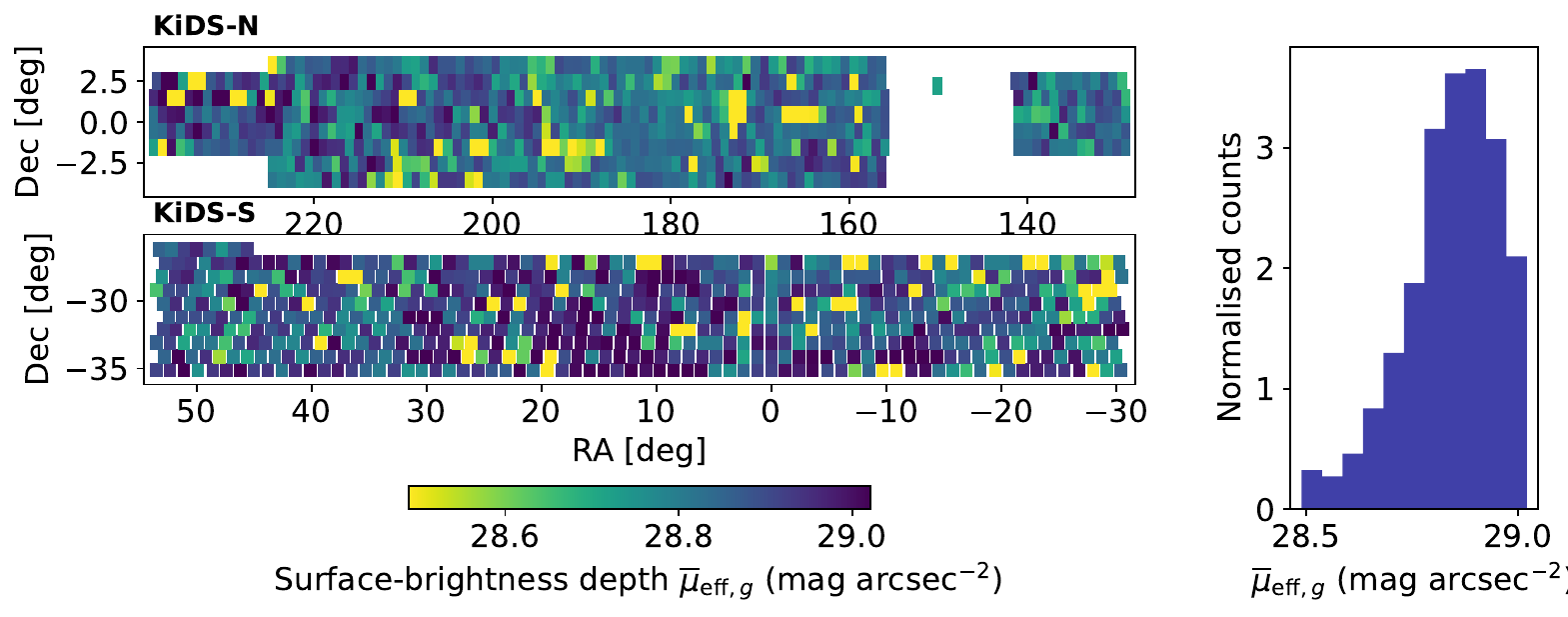}
        \caption{KiDS $g$-band surface-brightness depth distribution.}\label{kids_g_depth_map}
    \end{subfigure}

    % \vspace{1em}

    \begin{subfigure}{0.95\linewidth}
        \centering
        \includegraphics[width=\linewidth]{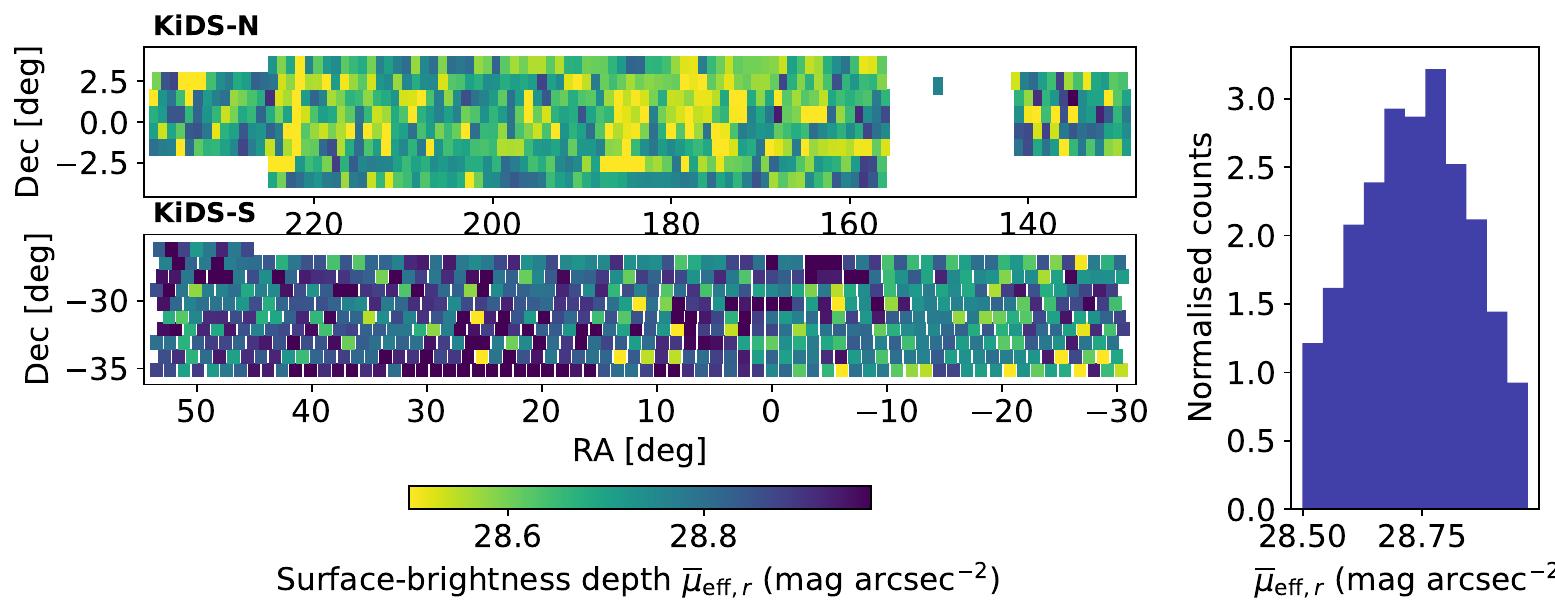}
        \caption{KiDS $r$-band surface-brightness depth distribution.}\label{kids_r_depth_map}
    \end{subfigure}

    % \vspace{1em}

    \begin{subfigure}{0.95\linewidth}
        \centering
        \includegraphics[width=\linewidth]{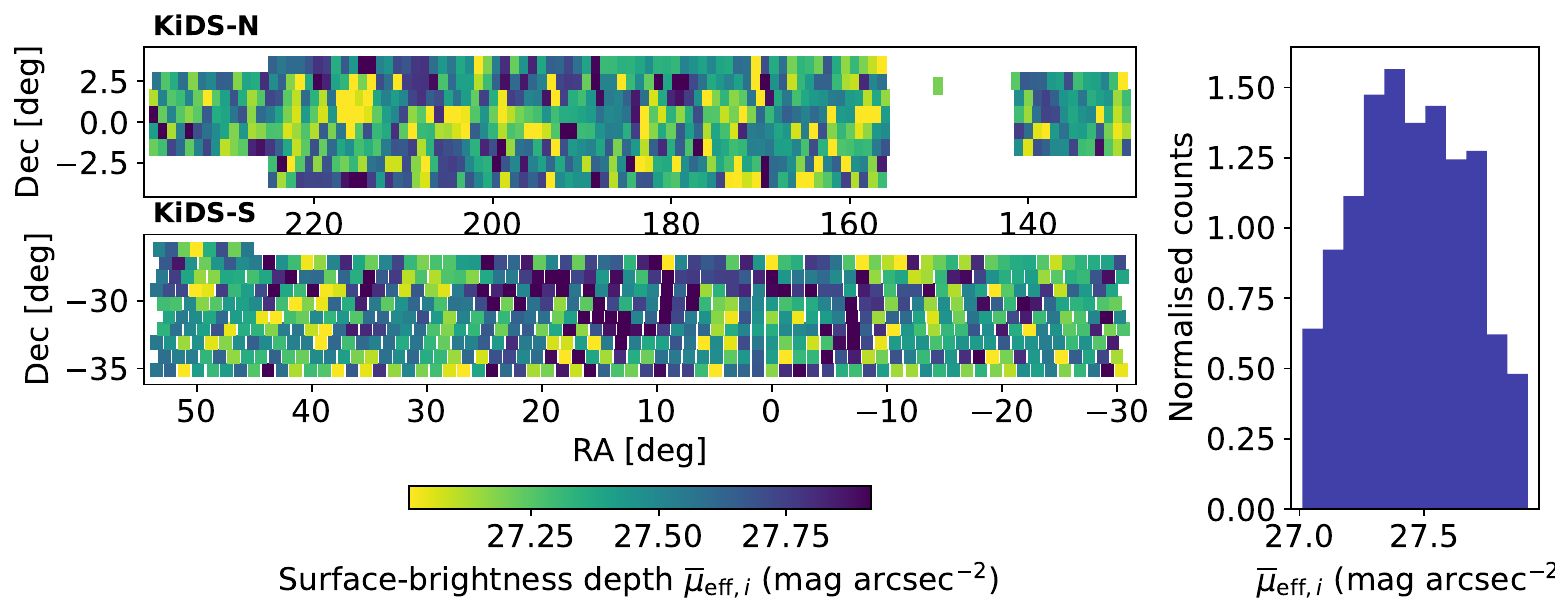}
        \caption{KiDS $i$-band surface-brightness depth distribution.}\label{kids_i_depth_map}
    \end{subfigure}

    \caption{KiDS surface-brightness depth maps in the $g$, $r$, and $i$ bands.}
    \label{fig:surface_brightness_depth}
\end{figure}

The $3\sigma$ surface-brightness detection limits in the KiDS $g$, $r$, and $i$ bands, measured over a $10\arcsec \times 10\arcsec$ region, for all KiDS tiles are shown in Fig. \ref{fig:surface_brightness_depth}.
For each tile in each band, the background rms ($\sigma$) noise was measured using \texttt{SEP}. For each tile, the $\sigma$ estimates were then used to compute the $3\sigma$ surface-brightness limit for a $10\arcsec \times 10\arcsec$ region using the relation
 \citep{Roman_surface_brightness_depth}:
\begin{equation}
    \mu_{lim}(3\sigma_{10\times10}) = -2.5\times log\left(\frac{3\sigma}{pix\times10}\right)+ZP+DMAG_x.
\end{equation}
Here, \textit{pix} is the pixel scale (0.2 arcsec/pix) of the KiDS data, and $ZP$ is the zero-point (0) of the KiDS data, and $DMAG_x$ adjusts for the systematic offset of ZP per band x and tile.

\FloatBarrier 

\section{Colour-colour diagram used for preselection}

\begin{figure}[h]
    \centering
    \includegraphics[width=\linewidth]{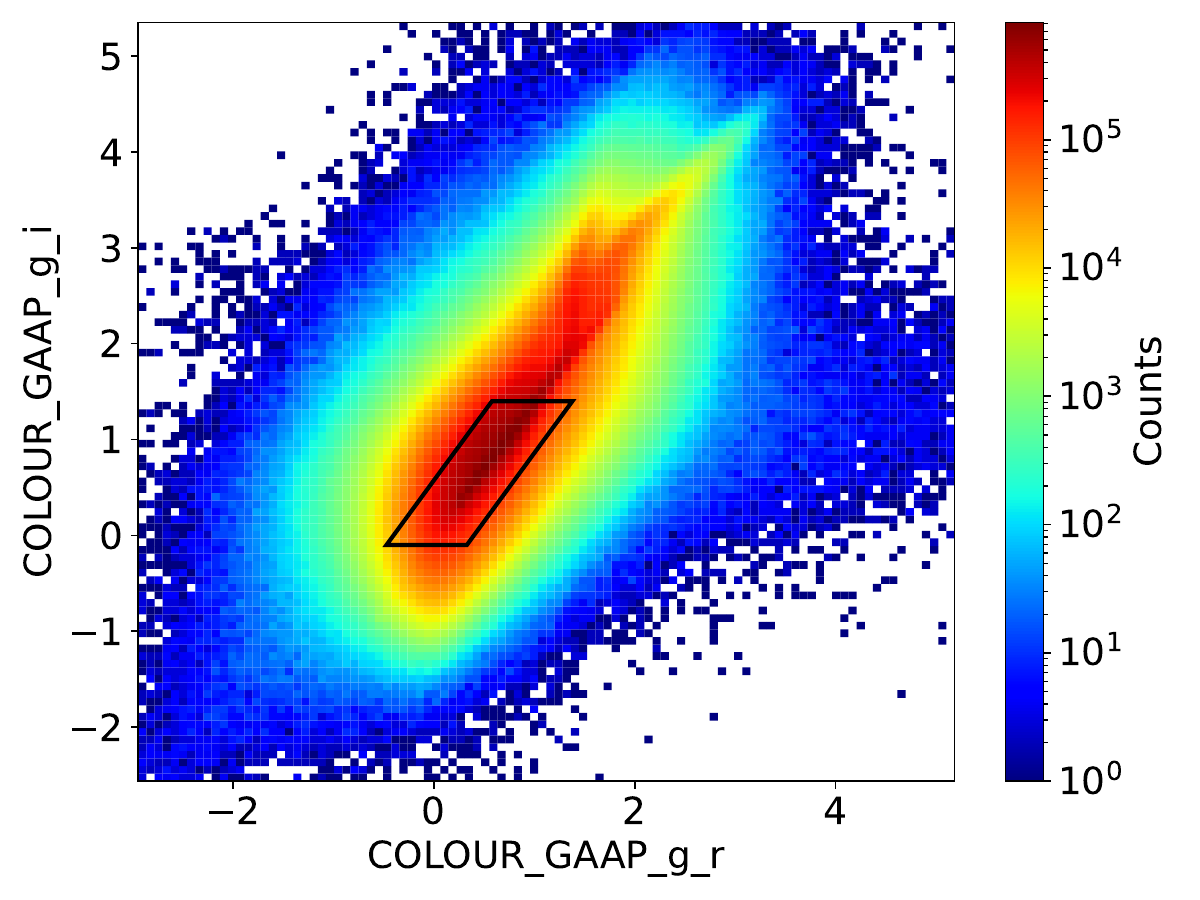}
    \caption{Colour-colour diagram of the KiDS DR5 parent sample using GAaP photometry. The ${\tt COLOUR\_GAAP\_g\_i}$ colour is constructed as ${\tt COLOUR\_GAAP\_g\_r} + {\tt COLOUR\_GAAP\_r\_i}$. The black box indicates the selection criteria used to preselect LSBG candidates. The adopted boundaries are designed to isolate galaxies with colours consistent with known LSBG populations, while minimising contamination from non-LSB sources.}
    \label{fig:color_preselctions}
\end{figure}

The colour-colour diagram of the full KiDS DR5 parent sample, together with the selection region used to preselect LSBG candidates, is shown in Fig.~\ref{fig:color_preselctions}. The adopted colour cuts, defined in terms of GAaP photometry, are consistent with the locus of LSBGs reported in previous studies \citep{Greco, Tanoglidis1, Zaritsky_SMUDGE, Su_LSBGNET}.

\section{Models and model performance}
\begin{figure*}
    \centering
    \begin{subfigure}{0.825\textwidth}
        \centering
        \includegraphics[width=\linewidth]{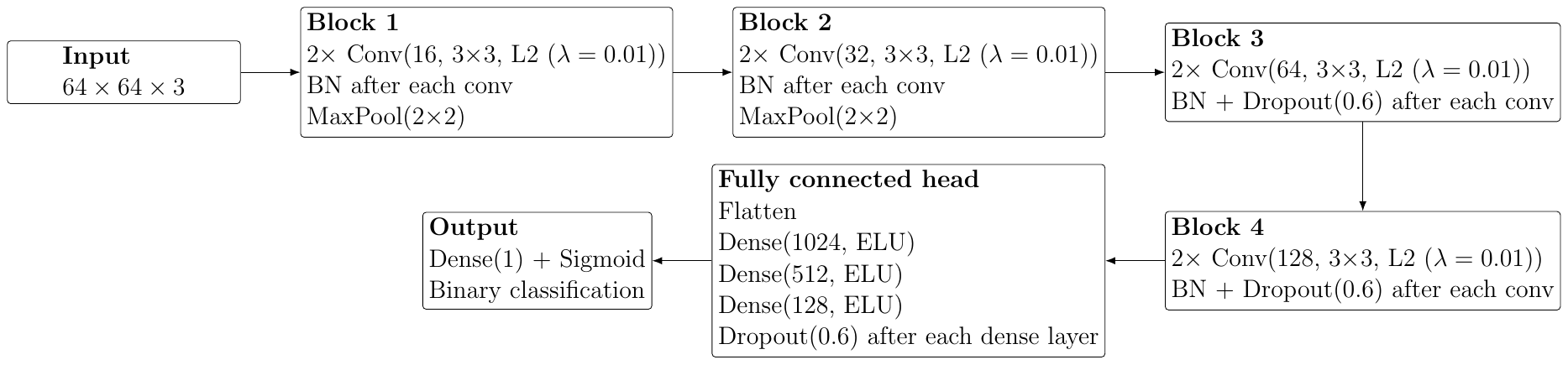}
        \caption{Architecture of the LSBG CNN model.}
        \label{fig:LSBG_CNN}
    \end{subfigure}

    \vspace{2mm}

    \begin{subfigure}{0.825\textwidth}
        \centering
        \includegraphics[width=\linewidth]{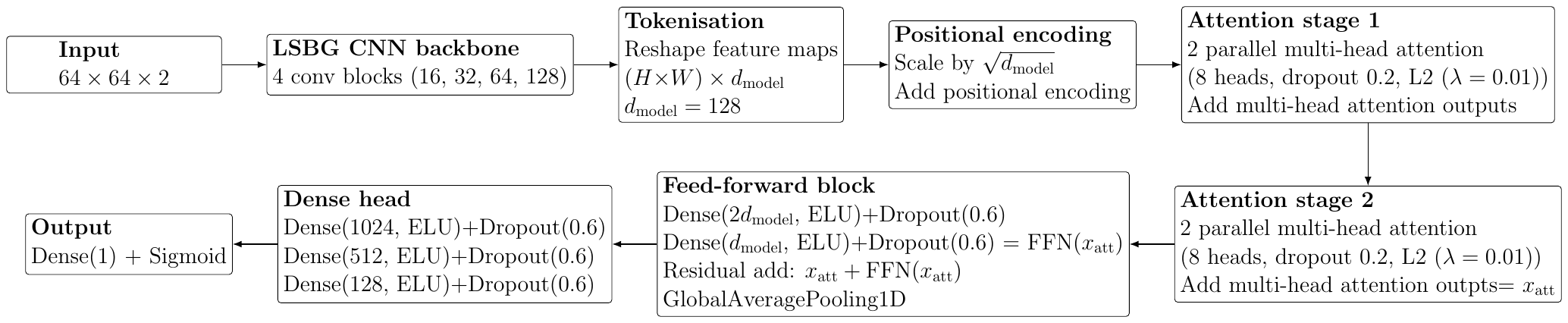}
        \caption{Architecture of the LSBG-DETR model.}
        \label{fig:lsbg_detr}
    \end{subfigure}

    \vspace{2mm}

    \begin{subfigure}{0.825\textwidth}
        \centering
        \includegraphics[width=\linewidth]{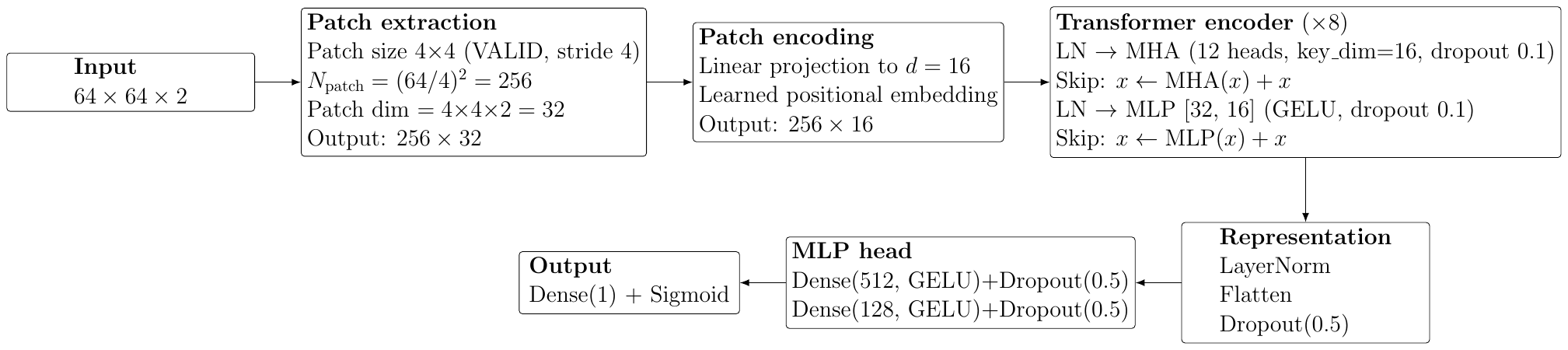}
        \caption{Architecture of the LSBG-VIT model.}
        \label{fig:lsbg_vit}
    \end{subfigure}

    \caption{
Architectures of the DL models used in this work.
Abbreviations used in the figure are as follows:
LN denotes layer normalisation;
BN denotes batch normalisation;
MHA denotes multi-head self-attention;
MLP denotes a multi-layer perceptron;
FFN denotes a feed-forward network;
ELU denotes the exponential linear unit activation;
GELU denotes the Gaussian error linear unit activation.
Residual (skip) connections indicate element-wise additions between block inputs and outputs. For further information about the models, we refer the reader to \citet{haressh_lsbs}.}
    \label{fig:model_architectures}
\end{figure*}

\begin{figure}[h]
    \centering
\includegraphics[width=0.95\columnwidth]{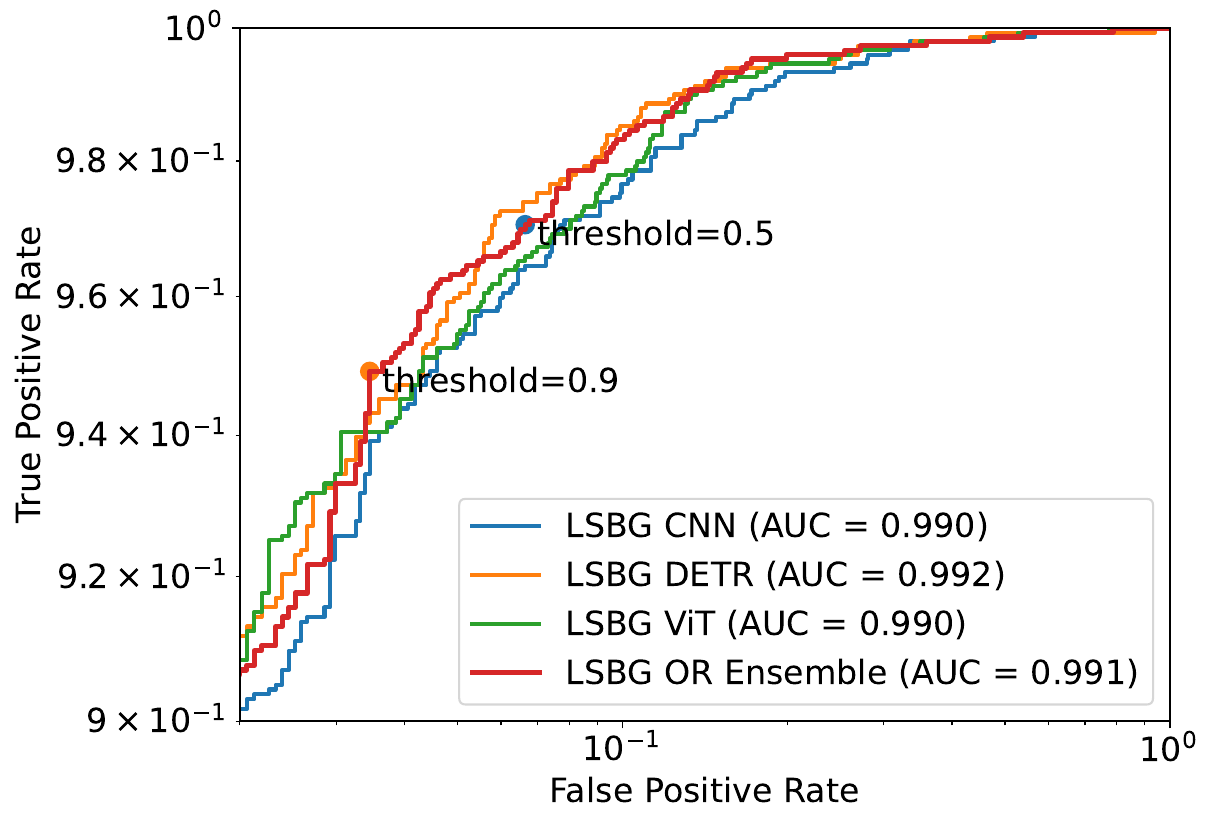}
    \caption{Receiver operating characteristic curve (ROC) of all the models presented in this work. The area under the ROC curve (AUC) for each model is given in the legend. The true positive rate and false positive rate corresponding to a detection threshold of 0.5 and 0.9 are marked by blue and orange dots.  }
    \label{fig:roc}
\end{figure}

The architectures of the DL models used in this work, namely the CNN, DETR, and ViT, are shown in Fig.~\ref{fig:model_architectures}. 
The receiver operating characteristic (ROC) curves of the three implemented models, together with the performance of the ensemble model, are shown in Fig.~\ref{fig:roc}. 
The code used to implement the models is publicly available at \url{https://github.com/hareesht23/LSBGs-in-DES-with-Transformers}.

\begin{figure}[h]
    \centering
    \includegraphics[width=0.9\linewidth]{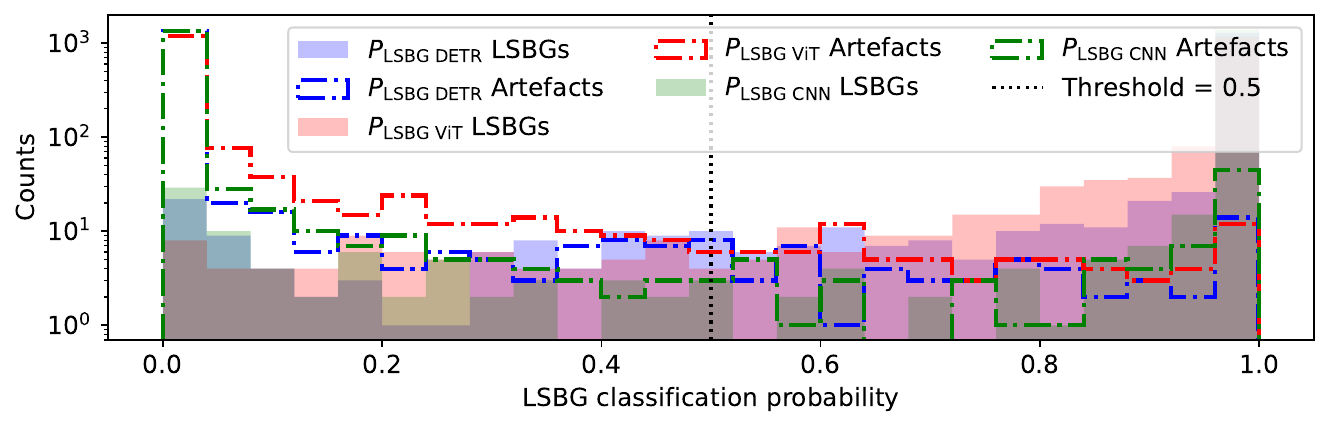}
\caption{Distribution of LSBG classification probabilities predicted by the three models: LSBG-DETR, LSBG-ViT, and LSBG-CNN. The vertical dashed line indicates the adopted selection threshold of $P = 0.5$.}    \label{fig:P_des_dist}
\end{figure}

\begin{figure}[h]
    \centering
    \includegraphics[width=0.9\linewidth]{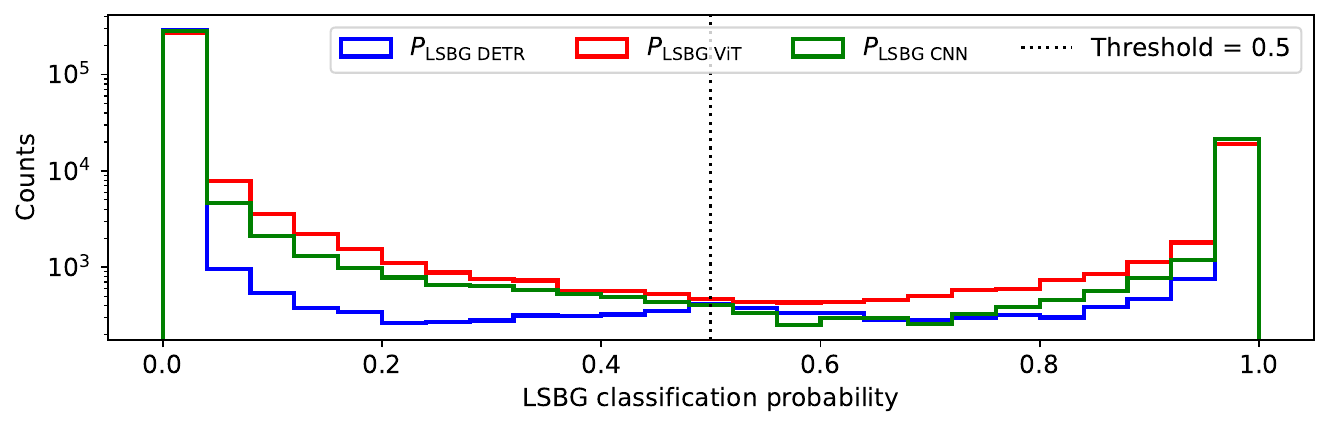}
\caption{Distribution of LSBG classification probabilities predicted by the three models: LSBG-DETR, LSBG-ViT, and LSBG-CNN. The vertical dashed line indicates the adopted selection threshold of $P = 0.5$.}    \label{fig:P_des_kids}
\end{figure}

\begin{figure}
    \centering
    \includegraphics[width=\linewidth]{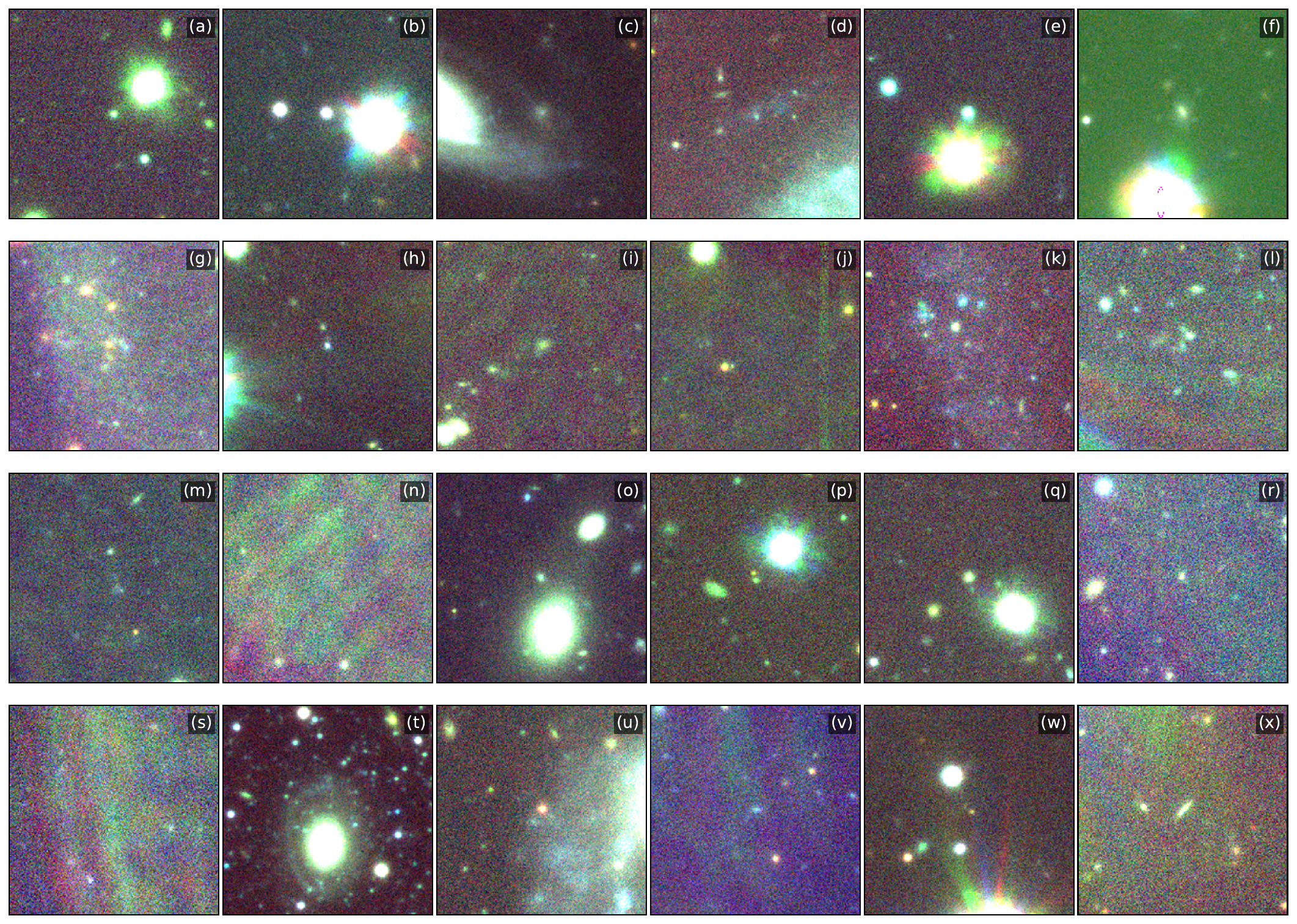}
    \caption{Examples of sources rejected by the DL models, i.e. objects for which all models assign probabilities $< 0.5$. The panels (a–x) illustrate a range of contaminants, including bright stars, compact galaxies, imaging artefacts, spiral arms of nearby galaxies, and noise-dominated detections that do not resemble typical LSBG morphology.}
    \label{fig:rejected_sources}
\end{figure}

The distribution of LSBG classification probabilities predicted by LSBG-DETR, LSBG-ViT, and LSBG-CNN, for the DES and KiDS datasets, is shown in Fig.~\ref{fig:P_des_dist} and Fig.~\ref{fig:P_des_kids}, respectively. Examples of sources having probabilities less than 0.5 in all the DL models are shown in Fig.~\ref{fig:rejected_sources}. 

\FloatBarrier
\onecolumn
\section{CIGALE input}
We present the parameters used for SED fitting in Table~\ref{TAB:CIGALE_input}.
\begin{table}[h]
\tiny
\centering
        \caption{Input parameters used in SED fitting with CIGALE.}
        \label{TAB:CIGALE_input}
        \footnotesize
        \begin{tabular}{l c}

\hline
Parameter & Values\\
\hline
\multicolumn{2}{c}{Delayed star formation history}\\
\hline
e-folding time of the main stellar population model (Myr) & {500, 1000, 1500,3000, 3500, 4500, 5500}\\
Age of the main stellar population& 300, 500, 800, 1000, 2500, \\
 in the galaxy (Myr) &3000, 5000, 6000, 8000, 9000, 10000, 12000\\
Mass fraction of the late burst population & 0\\
Normalise the SFH to produce one solar mass & True\\
\hline
\multicolumn{2}{c}{Stellar population synthesis  \cite{bruzal03}}\\
\hline
Initial mass function & \cite{chabrier03}\\
Metallicity & 0.008\\
Age of the separation between the young and the old star populations (Myr) & 10\\
\hline
\multicolumn{2}{c}{Dust attenuation \cite{CF2000}}\\
\hline
V-band attenuation in the interstellar medium & 0.001, 0.01, 0.05, 0.1, 0.15, 0.25, 0.3, 0.4\\
$\mu$ &  0.44, 0.5\\
Power law slope of the attenuation in the ISM & -0.7\\
Power law slope of the attenuation in the birth clouds & -0.7, -1, -1.3\\
\hline
\multicolumn{2}{c}{Redshifting (used only for estimating photo-$z$)}\\
\hline
Redshift & 2000 steps between 0.001 to 0.5\\
\hline
\end{tabular}
\end{table}  
\section{KiDS-LSBGs}
The description of the columns in the KiDS-LSBGs catalogue is shown in Table \ref{lsbs_table}.
\begin{table}
\centering
% \tiny
\caption{Description of the columns in the KiDS-LSBGs catalogue.}
\begin{tabular}{llp{13cm}}
\hline\hline
Column & Unit & Description \\
\hline
ID & -- & Internal galaxy identifier. \\

RA & deg & Right ascension (J2000). \\

Dec & deg & Declination (J2000). \\

$n$ & -- & S\'ersic index from the 2D S\'ersic fit to the galaxy surface brightness profile. \\

$\delta n$ & -- & Formal 1$\sigma$ uncertainty on the S\'ersic index. \\

$q$ & -- & Axis ratio $q = b/a$ of the S\'ersic model. \\

PA & deg & Position angle of the major axis of the S\'ersic model, measured from north ($0^\circ$) through east ($+90^\circ$). \\

$m_g$, $m_r$, $m_i$ & mag & Total apparent magnitudes in the \textit{g}-, \textit{r}-, and \textit{i}-bands, respectively, from the S\'ersic model fits. Corrected for Galactic extinction and photometric zero-point offsets. \\

$r_{\mathrm{eff},g}$, $r_{\mathrm{eff},r}$, $r_{\mathrm{eff},i}$ & arcsec & Non-circularised half-light radii in the \textit{g}-, \textit{r}-, and \textit{i}-bands, respectively. All $r_{\mathrm{eff}}$ values are measured along the semi-major axis.\\

$\delta r_{\mathrm{eff},g}$, $\delta r_{\mathrm{eff},r}$, $\delta r_{\mathrm{eff},i}$ & arcsec & Formal 1$\sigma$ uncertainties on the half-light radii in the \textit{g}-, \textit{r}-, and \textit{i}-bands. \\

$\overline{\mu}_{\mathrm{eff},g}$, $\overline{\mu}_{\mathrm{eff},r}$, $\overline{\mu}_{\mathrm{eff},i}$ & \magperarcsec{} & Mean surface brightness within $r_{\mathrm{eff}}$ in the \textit{g}-, \textit{r}-, and \textit{i}-bands, respectively. These values are corrected for inclination. \\

$\mu_{0,g}$, $\mu_{0,r}$, $\mu_{0,i}$ & \magperarcsec{} & Central surface brightness of the S\'ersic model in the \textit{g}-, \textit{r}-, and \textit{i}-bands, respectively. These values are corrected for inclination. \\

$\chi^2_{\mathrm{red,Ser}}$ & -- & Reduced $\chi^2$ of the S\'ersic model fit to the galaxy light profile. \\

BIC & -- & Bayesian Information Criterion of the S\'ersic model fit. \\

Nucleus & -- & Flag indicating the presence of an unresolved nuclear source (1 = nucleus detected, 0 = no nucleus). \\

$RA_{\mathrm{PSF}}$, $DEC_{\mathrm{PSF}}$ & deg & Best-fitting position of the unresolved nuclear PSF component. \\

$m_{g,\mathrm{PSF}}$, $m_{r,\mathrm{PSF}}$, $m_{i,\mathrm{PSF}}$ & mag & Apparent magnitudes of the unresolved nuclear PSF component in the \textit{g}-, \textit{r}-, and \textit{i}-bands, respectively. \\

$z$ & -- & Adopted galaxy redshift, obtained either from spectroscopy, cluster membership assignment, or photo-$z$ (see $z$ type). \\

Cluster & -- & Flag for cluster membership identifier (1 for cluster member). \\

$z$ type & -- &Origin of the adopted redshift. Values encode the source of $z$: \texttt{SED} indicates a photo-$z$ estimated from SED fitting; cluster-names are given for LSBGs assigned to a cluster; and spectroscopic survey names (2dFGRS, DESI, GAMA, SDSS, ALFALFA, FAST, GLADE) indicate $z$ obtained via redshift cross-matching to that spectroscopic catalogue. \\

Morphology & -- & Visual morphology classification: spiral (1), elliptical (2), featureless (-1), or ambiguous/uncertain morphology (-99). \\

$\overline{\mu}_{\mathrm{eff},r}^{z\text{-corr}}$ & \magperarcsec{} & Mean \textit{r}-band surface brightness within $r_{\mathrm{eff},r}$, corrected for cosmological surface-brightness dimming. -99, if the redshift is from SED fitting. \\

$\mathrm{SFR}_{\mathrm{SED}}$ & $M_\odot\,\mathrm{yr}^{-1}$ & Star-formation rate from SED fitting using the Bayesian star-formation history (SFH) model. \\

$\delta\mathrm{SFR}_{\mathrm{SED}}$ & $M_\odot\,\mathrm{yr}^{-1}$ & 1$\sigma$ uncertainty on the SED-derived star-formation rate. \\

$M_\star$ & $M_\odot$ & Stellar mass from Bayesian SED fitting. \\

$\delta M_\star$ & $M_\odot$ & 1$\sigma$ uncertainty on the SED-derived stellar mass. \\

$\chi^2_{\mathrm{SED}}$ & -- & Minimum reduced $\chi^2$ of the best-fitting SED model from CIGALE. \\

\hline
\end{tabular}
\label{lsbs_table}
\end{table}

\FloatBarrier 
% ------------------------------------------------------------------------------

\section{Non-LSBGs$_z$}
The description of the columns in the Non-LSBGs$_z$ catalogue is shown in Table \ref{non_lsbs_table}.
\begin{table}
% \small
\centering
\caption{Description of the columns in the Non-LSBGs$_z$ catalogue.}
\begin{tabular}{llp{13cm}}
\hline\hline
Column & Unit & Description \\
\hline
ID & -- & Internal galaxy identifier. \\

RA & deg & Right ascension (J2000). \\

Dec & deg & Declination (J2000). \\

$n$ & -- & S\'ersic index from the 2D S\'ersic fit to the galaxy surface brightness profile. \\

$\delta n$ & -- & Formal 1$\sigma$ uncertainty on the S\'ersic index. \\

$q$ & -- & Axis ratio $q = b/a$ of the S\'ersic model. \\

PA & deg & Position angle of the major axis of the S\'ersic model, measured from north ($0^\circ$) through east ($+90^\circ$). \\

$m_g$, $m_r$, $m_i$ & mag & Total apparent magnitudes in the \textit{g}-, \textit{r}-, and \textit{i}-bands, respectively, from the S\'ersic model fits. Corrected for Galactic extinction and photometric zero-point offsets. \\

$r_{\mathrm{eff},g}$, $r_{\mathrm{eff},r}$, $r_{\mathrm{eff},i}$ & arcsec & Non-circularised half-light radii in the \textit{g}-, \textit{r}-, and \textit{i}-bands, respectively. All $r_{\mathrm{eff}}$ values are measured along the semi-major axis. \\

$\delta r_{\mathrm{eff},g}$, $\delta r_{\mathrm{eff},r}$, $\delta r_{\mathrm{eff},i}$ & arcsec & Formal 1$\sigma$ uncertainties on the half-light radii in the \textit{g}-, \textit{r}-, and \textit{i}-bands. \\

$\overline{\mu}_{\mathrm{eff},g}$, $\overline{\mu}_{\mathrm{eff},r}$, $\overline{\mu}_{\mathrm{eff},i}$ & \magperarcsec{} & Mean surface brightness within $r_{\mathrm{eff}}$ in the \textit{g}-, \textit{r}-, and \textit{i}-bands, respectively. These values are corrected for inclination. \\

$\mu_{0,g}$, $\mu_{0,r}$, $\mu_{0,i}$ & \magperarcsec{} & Central surface brightness of the S\'ersic model in the \textit{g}-, \textit{r}-, and \textit{i}-bands, respectively. These values are corrected for inclination. \\

$\chi^2_{\mathrm{red,Ser}}$ & -- & Reduced $\chi^2$ of the S\'ersic model fit to the galaxy light profile. \\

BIC & -- & Bayesian Information Criterion of the S\'ersic model fit. \\

Nucleus & -- & Flag indicating the presence of an unresolved nuclear source (1 = nucleus detected, 0 = no nucleus). \\

$RA_{\mathrm{PSF}}$, $DEC_{\mathrm{PSF}}$ & deg & Best-fitting position of the unresolved nuclear PSF component. \\

$m_{g,\mathrm{PSF}}$, $m_{r,\mathrm{PSF}}$, $m_{i,\mathrm{PSF}}$ & mag & Apparent magnitudes of the unresolved nuclear PSF component in the \textit{g}-, \textit{r}-, and \textit{i}-bands, respectively. \\

$z$ & -- & Adopted galaxy redshift, obtained either from spectroscopy, cluster membership assignment, or photo-$z$ (see $z$ type). \\

Cluster & -- & Flag for cluster membership identifier (1 for cluster member). \\

$z$ type & -- & Origin of the adopted redshift.  \\

Morphology & --  & Visual morphology classification: spiral (1), elliptical (2), featureless (-1), or ambiguous/uncertain morphology (-99) \\

$\overline{\mu}_{\mathrm{eff},r}^{z\text{-corr}}$ & \magperarcsec{} & Mean \textit{r}-band surface brightness within $r_{\mathrm{eff},r}$, corrected for cosmological surface-brightness dimming to the adopted reference redshift. \\

\hline
\end{tabular}
\label{non_lsbs_table}
\end{table}
\section{Quenched fractions across environments and redshift }
The quenched fractions ($f_q$) of non-UDGs and UDGs in different environments, split into different redshift bins, are shown in Table \ref{tab:quenched_fractions}.
\begin{table}
\centering
\small
\setlength{\tabcolsep}{3pt}
\renewcommand{\arraystretch}{1.15}
\caption{Quenched fractions ($f_q$) of non-UDGs and UDGs in different environments, defined as $\Delta{\rm MS}<-0.4$ dex.
Quenched fractions are reported as $f_q$~[\%], with the total number of galaxies in parentheses.
Median stellar masses are given in $\log_{10}(M_\star/M_\odot)$.}
\begin{tabular}{lcccc}
\hline
\multicolumn{1}{c}{%
\parbox[c]{2.8cm}{\centering Redshift bin \&\\
Environment}
} &
\multicolumn{2}{c}{$f_q$ [\%] $(N)$} &
\multicolumn{2}{c}{$\log_{10}(M_\star/M_\odot)$} \\
 & non-UDG & UDG & non-UDG & UDG \\
\hline
\multicolumn{5}{l}{$0.00 < z \le 0.03$} \\
All     & 40.5 (1453) & 54.4 (158) & 7.62 & 7.78 \\
Non-cluster   & 11.3 (768)  & 7.5  (40)  & 7.99 & 7.79 \\
Cluster & 73.1 (685)  & 70.1 (118) & 6.70 & 7.77 \\
\hline
\multicolumn{5}{l}{$0.03 < z \le 0.06$} \\
All     & 7.9  (1155) & 36.4 (173) & 8.61 & 8.10 \\
Non-cluster   & 4.2  (1067) & 4.6  (87)  & 8.61 & 8.17 \\
Cluster & 52.3 (88)   & 68.6 (86)  & 8.60 & 8.01 \\
\hline
\multicolumn{5}{l}{$0.06 < z \le 0.10$} \\
All     & 3.7  (1322) & 21.0 (81)  & 9.05 & 8.59 \\
Non-cluster   & 2.5  (1271) & 1.9  (52)  & 9.05 & 8.58 \\
Cluster & 33.3 (51)   & 55.2 (29)  & 9.04 & 8.61 \\
\hline
\multicolumn{5}{l}{$0.10 < z \le 0.20$} \\
All     & 1.3  (548)  & 9.1  (22)  & 9.41 & 8.85 \\
Non-cluster   & 1.3  (537)  & 5.0  (20)  & 9.41 & 8.85 \\
Cluster & 0.0  (11)   & 50.0 (2)   & 9.17 & 8.76 \\
\hline
\end{tabular}
\label{tab:quenched_fractions}
\end{table}

\end{appendix}
\end{document}